\documentclass[12pt]{iopart}

\usepackage{graphicx}
\usepackage{epsfig}
\usepackage{verbatim}

\usepackage[T1]{fontenc}
\usepackage{eufrak}

\begin{document}

\newcommand{\be}{\begin{equation}}
\newcommand{\ee}{\end{equation}}
\newcommand{\beqn}{\begin{eqnarray}}
\newcommand{\eeqn}{\end{eqnarray}}
\newcommand{\dfrac}{\frac}

\title{Non-equilibrium quantum dynamics after local quenches}

\author{Uma Divakaran $^1$, Ferenc Igl\'oi $^{2,3}$, Heiko Rieger $^1$}

\address{$^1$ Theoretische Physik, Universit\"at des Saarlandes, 
66041 Saarbr\"ucken, Germany\\
$^2$ Research Institute for Solid State Physics and Optics,
H-1525 Budapest, P.O.Box 49, Hungary \\
$^3$ Institute of Theoretical Physics,
Szeged University, H-6720 Szeged, Hungary
}

\begin{abstract}
We study the quantum dynamics resulting from preparing a
one-dimensional quantum system in the ground state of initially two
decoupled parts which are then joined together (local quench).
Specifically we focus on the transverse Ising chain and compute the
time-dependence of the magnetization profile, $m_l(t)$, and
correlation functions at the critical point, in the ferromagnetically
ordered phase and in the paramagnetic phase. At the critical point
we find finite size scaling forms for the nonequilibrium magnetization
and compare predictions of conformal field theory with our numerical
results. In the ferromagnetic phase the magnetization profiles 
are well matched by our predictions from a quasi-classical calculation.
\end{abstract}


\maketitle

\section{Introduction}

Recently there has been an increased interest in nonequilibrium
relaxation processes in closed quantum systems following a sudden
change of the parameters of the Hamiltonian (quantum quench)
\cite{rig-06,rig-07,caz-06,cc-06,cc-07,sotiriadis_cardy,l-07,gdlp-07,kol-08,bs-08,qr-08,ke-08,
fcmse-08,cfmse-08,bpgda-09,ro-09,fm-09,rossini,igloi_rieger2,cef-11}.
Theoretically during the initial period ($t<0$) the system is described
by a Hamiltonian ${\cal H}_0$ with ground state
$|\Psi_0\rangle$, which is suddenly changed to a new Hamiltonian,
${\cal H}$, for time $t \ge 0$.  The new state of the system - in the
Schr\"odinger picture - is time-dependent:
\be
|\Psi(t)\rangle=\exp(-it{\cal H})|\Psi_0\rangle\;,
\label{psi_t}
\ee
and so is with an operator, say ${\cal O}$, which is expressed at time $t>0$ in the Heisenberg picture as:
\be
{\cal O}(t)=\exp(-it{\cal H}){\cal O}\exp(it{\cal H})\;.
\label{O_t}
\ee
Generally one is interested in the relaxation
of the local order-parameter,
\be
m_r(t)=\langle \Psi_0|{\cal O}_r(t)|\Psi_0\rangle\;,
\label{m_r}
\ee
at site, $r$ or the time-dependence of the
autocorrelation function: 
\be
G_r(t,t')=\langle \Psi_0|{\cal O}_r(t){\cal O}_r(t')|\Psi_0\rangle\;.
\label{G_r}
\ee
After long times, $t \gg 1$, the system is expected to relax to a stationary state,
in which one measures the correlation function: 
\be
C_t(r_1,r_2)=\langle \Psi_0|{\cal O}_{r_1}(t){\cal O}_{r_2}(t)|\Psi_0\rangle\;.
\label{C_t}
\ee
Most often one considers global quenches, when the parameters are modified uniformly at all points of the
sample. Experimentally this process can be realized with ultracold atomic gases in optical lattices \cite{BE_exp,spinor,1d,1d_dyn}.
After a global quench a quantum system is expected to relax to a thermal (or thermal-like) state, such
that the local magnetization (as well as the autocorrelation function) vanishes exponentially,
$m_r(t) \sim \exp(-t/\tau)$, with a relaxation (or phase coherence) time $\tau$. Similarly, the correlation function in the stationary state behaves as:
\be
C_t(r_1,r_2) \sim \exp(-|r_1-r_2|/\xi)\;,
\label{xi}
\ee
with a finite non-equilibrium correlation length, $\xi$.
This thermalization of the system is generally explained in terms of quasi-particles \cite{cc-07,sachdev_young}, which are emitted
during the quench
at each point of the sample and which travel with a constant velocity, $v$. Quasi-particles which are originated
at nearby sites (which are within the correlation length) are quantum entangled, but other quasi-particles are incoherent.
Incoherent particles reaching a reference point, $r$, will result in the reduction of the value of a local observable and this
process is responsible for the exponential decay of the (auto)correlation function. In a system with boundaries
these quasi-particles are reflected at the boundaries, which results in more complicated time-dependence, among
others reconstruction of the local magnetization \cite{igloi_rieger2}.

Besides global quenches one also considers local quenches, when the parameters are modified only at a
restricted region of the system \cite{ep-07,cc-07loc}. Experimentally this type of problem can be realized in the x-ray
absorption problem in metals \cite{x-ray}, where the creation of a hole plays the role of a local defect and when a conduction
electron fills the hole this potential is suddenly switched off.
Nonequilibrium dynamics following a local quench has been first studied in the context of
the entanglement entropy, ${\cal S}$, in critical quantum spin chains \cite{ep-07,cc-07loc,ekpp-08}.
If we consider two disconnected half
chains for time $t<0$, which are suddenly joint together for $t \ge 0$, then the time-evolution of the
entanglement entropy of the two half chains is found to evolve in time asymptotically as:
\be
{\cal S}(t)=\dfrac{c}{3}\log t + c_1\;,
\label{S_t}
\ee
where $c$ is the central charge of the Virasoro algebra and $c_1$ is a non-universal constant.
In view of Eq.(\ref{S_t}) basic informations about the critical properties of quantum spin chains, such as the value of the central charge, can be obtained through nonequilibrium local quench dynamics.
The result in Eq.(\ref{S_t}), which has been first obtained numerically for a free-fermion model \cite{ep-07}, has been
derived through conformal invariance \cite{cc-07loc}. It has also been generalized for different positions of the defect \cite{ep-07,cc-07loc}, as well as
for varying strength of the defect in quantum Ising and XX-chains \cite{iszl-09}. For a large finite
chain of total length, $L$, the appropriate expression is conjectured to be \cite{iszl-09}:
\be
{\cal S}_L(t)=\frac{c}{3} \log \left| \frac{L}{ \pi} \sin \frac{ \pi t}{L} \right|+c_1\;.
\label{S_t_conf1}
\ee
The conformal mapping, which has been used for the calculation of the entanglement entropy can be applied \cite{cc-07loc}
to study the time-dependence of one-point functions (such as the local magnetization in Eq.(\ref{m_r})), as well as two-point
functions (such as the correlation function in Eq.(\ref{C_t})) in a critical quantum spin chain. In contrary
to global quenches these results indicate a power-low relaxation as well as power-low type correlations
in the stationary state. This is understandable within the quasi-particle picture, since at a local quench
the quasi-particles are expected to be emitted only at a restricted region of the sample and therefore these are
quantum entangled. If these quasi-particles arrive after time $t$ to different positions of the chain, say $r_1$ and $r_2$,
will result in correlations.

Besides the conformal results which are mentioned above several problems connected with local quench dynamics
are still unexplored. To our knowledge there are no numerical investigations, which could be used to check
the validity of the conformal conjectures. There are no results about the autocorrelation function and no
information is known about nonequilibrium relaxation following a local quench to the ordered or to the disordered
phase of a quantum spin chain. In this paper we aim to fill this gap and perform detailed numerical investigations
about these open questions. As a model we use the transverse Ising spin chain, which is a prototypical quantum system
having ordered and disordered phases as well as a quantum critical point \cite{pfeuty}. Using free-fermionic
techniques \cite{lieb,barouch_mccoy,pfeuty}
we have studied numerically the nonequilibrium relaxation processes for the magnetization, the autocorrelation
and the equal-time correlation function in large finite systems. Most of our investigations are performed at
the quantum critical point, so that we could compare our results with the conformal predictions. However, we also
studied quenches to the ordered and to the disordered phases and these results
are then compared with semi-classical calculations..

The structure of the paper is the following. The model and the numerical method of the calculation is
introduced in Sec.\ref{sec:model}. Results of the calculations at the critical point, in the ferromagnetic and in the
paramagnetic phases are presented in Sections \ref{sec:critical}, \ref{sec:ferro} and \ref{sec:para}, respectively.
Our conclusions are summarized in Sec.\ref{sec:concl}.
\section{Model}
\label{sec:model}
The system we consider in this paper is the transverse Ising chain
of finite length $L$ with open boundaries defined by the Hamiltonian:
\be
{\cal H}=-\dfrac{1}{2}\left[ \sum_{l=1}^{L-1} 
J_l\sigma_l^x \sigma_{l+1}^x +\sum_{l=1}^{L} h_l \sigma_l^z\right] \;,
\label{hamilton}
\ee
in terms of the Pauli-matrices $\sigma_l^{x,z}$ at site $l$
and bond strengths $J_l$, which are all equal $J_l=1$ except 
the central bond $J_{L/2}$, which is $J_{L/2}=0$ for time $t\le0$. Similarly, the transverse fields are homogeneous, $h_l=h$,
except at the central sites, which are $h_{L/2}=h_{L/2+1}=0$ for time $t\le0$ for fixed-spin
boundary conditions. (For free-spin boundary conditions these are $h_{L/2}=h_{L/2+1}=h$ for time $t\le0$, too.)
Generally we measure distances from the defect and use the variable:
\be
r=l-\dfrac{L}{2}\;.
\label{r}
\ee
The order-parameter operator of the system is, $\sigma_l^x$,
what should be inserted in the formulae of the autocorrelation and the correlation functions, see in
Eqs.(\ref{G_r}) and (\ref{C_t}), respectively. For the local magnetization in Eq.(\ref{m_r}) one
should have: $m_r(t)=\lim_{b\to0_+}\,_b\langle \Psi_0 | \sigma_r^x(t) |
\Psi_0 \rangle_b$, where $|\Psi_0\rangle_b$ is the
ground state of the initial Hamiltonian (\ref{hamilton}) in the
presence of an external longitudinal field $b$.  According to
 \cite{yang} this can be written as the off-diagonal matrix-element of
the Hamiltonian (\ref{hamilton}):
\be
m_r(t)=\langle \Psi_0 | \sigma_r^x(t) | \Psi_1 \rangle\;,
\label{mag}
\ee
where $|\Psi_1\rangle$ is the first excited state
of the initial Hamiltonian ($t<0$). For fixed-spin boundary condition the ground-state
is exactly degenerate with $|\Psi_1\rangle$.
In the initial state and in the thermodynamic limit, $L \to \infty$, there is spontaneous
ferromagnetic order in the system, $m_r(0)>0$, for $h<h_c=1$. On the contrary, for stronger
transverse fields, $h>h_c$, the magnetization is vanishing. At the quantum critical point
the magnetization vanishes as a power-law: $m_r(0) \sim L^{-x}$, for bulk spins: $r=O(1)$,
and $m_{\pm L/2}(0) \sim L^{-x_s}$, for surface spins. Here the magnetization exponent $x$ and the surface magnetization
exponent $x_s$ is known exactly: $x=1/8$ and $x_s=1/2$.

The Hamiltonian in Eq.(\ref{hamilton}) can be expressed in terms of
free fermions  \cite{lieb,barouch_mccoy,pfeuty}, which has been used in
previous studies of its non-equilibrium properties
 \cite{igloi_rieger,rossini,igloi_rieger2}. The magnetization profile, as well
as the (auto)correlation functions can be expressed in terms of Pfaffians, which
are then evaluated as the square-root of an antisymmetric matrix, which has a rank $O(L)$.
In the following Sections we use these techniques to calculate different nonequilibrium quantities.

\section{Local quenches at the critical point}
\label{sec:critical}
\subsection{Conformal field theory - a reminder}
\label{sec:conf}
\begin{figure}[h!]
\begin{center}
\includegraphics[width=8cm,angle=0]{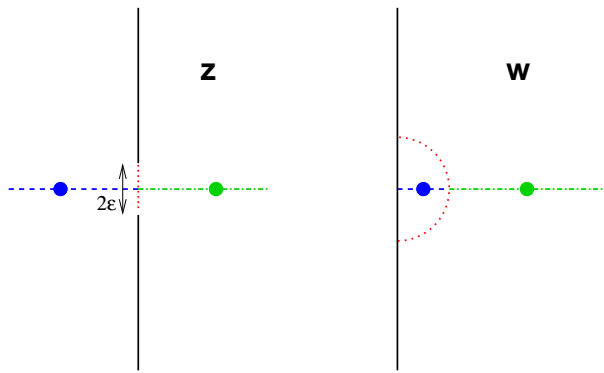}
\caption{
\label{fig:conf} (Color online)
Space-time region $z=r+\imath \tau$ with two cuts starting at $\pm \imath \epsilon$
(left) which is mapped to the half-plane $w$ through
the conformal transformation in Eq.(\ref{Joukowsky}). At the end of the calculation one should take $\tau \to \imath t$.}
\end{center}
\end{figure}

Here we recapitulate the basic results by Calabrese and Cardy \cite{cc-07loc} about the use
of conformal field theory for local quenches at the critical point. The system is
represented in a space-time region with coordinates: $(r,t)$ and the two
parts of the chain are decoupled for $t<0$ and these are joined at $t=0$ and
we measure the properties of the system at $t>0$. In the path integral formalism one introduces damping factors:
$\exp(-\epsilon{\cal H})$, so that in Eq.(\ref{O_t}) we have
$\exp(-\imath t{\cal H}) \to \exp(-\imath t{\cal H}-\epsilon{\cal H})$ and
$\exp(\imath t{\cal H}) \to \exp(\imath t{\cal H}-\epsilon{\cal H})$. For computational
simplicity the operators are inserted at imaginary times: $\tau=\imath t$ and one works in the complex plane
$z=r+\imath \tau$. Here, due to the local quench we have two cuts starting at $\pm \imath \epsilon$ and ending at
$\pm \imath \infty$. This is represented in the left panel of Fig.\ref{fig:conf}.

This geometry is mapped by the Joukowsky transformation:
\be
w=\dfrac{z}{\epsilon}+\sqrt{\left( \dfrac{z}{\epsilon}\right)^2+1},\quad z=\epsilon\dfrac{w^2-1}{2w}\;,
\label{Joukowsky}
\ee
into the half plane, ${\rm Re}~w >0$. At the boundaries of the slits in the $z$-plane as well as
at the surface of the $w$-plane one should impose boundary conditions, which can be either free or fixed spin
boundary conditions. In the $w$-plane the asymptotic form of the one- and two-point functions are generally
known due to conformal invariance \cite{cardy-84,cardy-90,bx-91,ti-97}. These results are then transformed back to the $z$ plane and at the end
of the calculation one analytically continues the final result as $\tau \to \imath t$.

We note that the conformal results obtained in this way are valid for infinitely long chains, i.e. for $L \to \infty$.

\subsection{Magnetization}
\label{sec:magn}
We have calculated the relaxation of the magnetization in the transverse-field Ising chain following a
local quench starting with two different type of initial state. First, we consider an initial state
with fixed spins at the defect, in which case the initial defect magnetization stays finite. In this
type of setting one can make a direct comparison with the conjectures of conformal invariance. In the
second type of calculation in the initial state we use free boundary conditions at the defect. Then
we calculate the time-dependence of the off-diagonal order-parameter in Eq.(\ref{mag}), which has a vanishing
initial value at the defect in the thermodynamical limit.

\subsubsection{Fixed-spin boundary condition}
\label{sec:magn_fix}

In this section in the initial state, $t<0$,
the two half chains are prepared with fixed spins at the boundaries:
$\langle \sigma^x_{L/2} \rangle=\langle \sigma^x_{L/2+1} \rangle=1$, which is obtained by fixing $h_{L/2}=h_{L/2+1}=0$
in the Hamiltonian in Eq.(\ref{hamilton}). We use the superscript, $^{(+)}$, to refer for fixed-spin boundary condition.
The magnetization profile in the initial state is known from conformal invariance \cite{cardy-90,bx-91}:
\be
m_r^{(+)}(t=0,L) \sim \left| \dfrac{L}{2\pi} \sin \pi \dfrac{2r}{L}\right|^{-x} \left( \cos \pi \dfrac{r}{L}\right) ^{x_s}\;,
\label{m+0}
\ee
which behaves close to the defect: $m_r^{(+)}(t=0) = A|r|^{-x}$, $|r| \ll L/2$. This is the well-known
result by Fisher and de Gennes \cite{FdeG}.

After the local quench, for $t>0$, we take $h_{L/2}=h_{L/2+1}=1$ (and $J_{L/2}=1$) and study the evolution of the
magnetization. In the limit $|r| \ll L/2$ the conformal mapping in the Sec.\ref{sec:conf} leads to the result \cite{cc-07loc}:
\be
m_r^{(+)}(t)=\left\{
\begin{array}{ll}
 A|r|^{-x} & t<r,\\
 A\left( \dfrac{\epsilon}{t^2-r^2}\right)^x & t>r\;.
\label{m_conf}
\end{array}
\right.
\ee
This can be interpreted, that for $t<r$ the magnetization keeps its initial value until the quasiparticles
from the defect arrive at $t=r$ and afterwards for $r \ll t \ll L/2$ the decay is given by $m_r^{(+)}(t)\sim t^{-2x}$.
This decay involves twice of the magnetization exponent and this behaviour is similar to that of
the equilibrium autocorrelation function at the critical point.

\begin{figure}[h!]
\begin{center}
\includegraphics[width=8cm,angle=0]{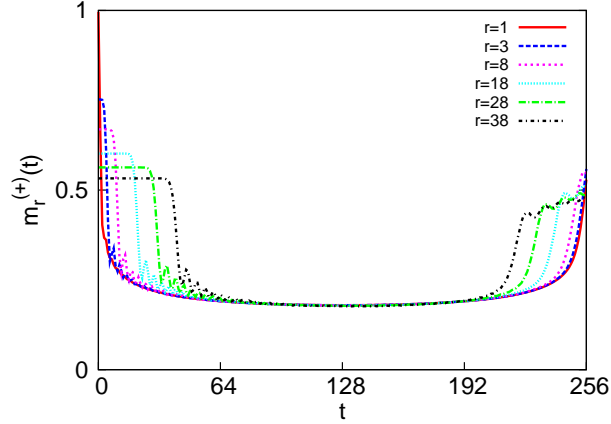}
\caption{
\label{fig:fixed_r} (Color online)
Temporal evolution of the magnetization at the critical point after a local quench having fixed spins at the defect
in the initial state in a finite system of length $L=256$.}
\end{center}
\end{figure}

In order to check the conformal result in Eq.(\ref{m_conf}) we have calculated the time-dependence of the
magnetization in finite chains of length up to $L=256$. For the largest chain and for different values of $r$
the relaxation of the magnetization is shown in Fig.\ref{fig:fixed_r}. In agreement with the quasi-particle picture and with
the results of conformal invariance the local magnetization stays unchanged until $t<r$, which is followed
by a fast decrease. Due to the finite size of the system the magnetization has a finite, $L$-dependent
limiting value and for $t>L/2$ the magnetization starts to increase.

\begin{figure}[h!]
\begin{center}
\includegraphics[width=8cm,angle=0]{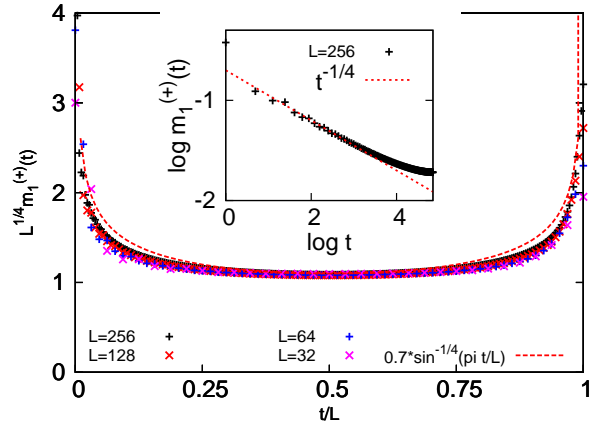}
\caption{
\label{magfixed0} (Color online)
Temporal evolution of the magnetization at the defect after 
a local quench from a fixed spin at the critical point. In the main
panel the scaled profiles are shown for different values of $L$, where the curve with the dashed line is given in Eq.(\ref{m1+scale}).
In the inset the log-log plot of $m_1^{(+)}(t,L)$ is shown for $L=256$, where the conformal prediction
for $L \to \infty$ is given by the dashed straight line.}
\end{center}
\end{figure}

We have studied in more detail the behavior of the magnetization at the defect:
$l=L/2$, i.e. for $r=1$. For short times, $1 \gg t \gg L/2$, the decay is compatible with the conformal
prediction: $m_1^{(+)}(t)\sim t^{-1/4}$ is seen in the inset of Fig.\ref{magfixed0} in which
the magnetization is shown as a function of time in a log-log plot. For longer time the boundaries of the chain
start to influence the relaxation and the critical defect magnetization is expected
to satisfy the scaling behavior: $m_1^{(+)}(t,L)=b^{-2x}m_1^{(+)}(t/b^z,L/b)$, when lengths are rescaled by a factor
$b>1$. In this relation the dynamical exponent is $z=1$ and by taking the scaling factor as $b=t$ in the limit $L \gg t$
we recover the conformal result: $m_1^{(+)}(t)\sim t^{-2x}$. It is more interesting to take $b=L$, when one obtains:
$m_1^{(+)}(t,L)=L^{-2x}{\mu}_1^{(+)}(t/L)$. Here the scaling function ${\mu}_1^{(+)}(\tau)$ for small $\tau=t/L$ behaves as: ${\mu}_0^{(+)}(\tau) \sim \tau^{-2x}$. Numerical results for the scaling function
for different values of $L$ are shown in the main panel of Fig.\ref{magfixed0}, which is found to be
well approximated by the function $B \left[ \sin(\pi \tau)\right]^{-2x}$, thus we have the conjecture:
\be
m_1^{(+)}(t,L)\propto L^{-2x}\left[ \sin\left(\pi \dfrac{t}{L}\right)\right]^{-2x}\;.
\label{m1+scale}
\ee

\subsubsection{Free-spin boundary condition}
\label{sec:magn_free}

\begin{figure}[t]
\begin{center}
\includegraphics[width=12cm,angle=0]{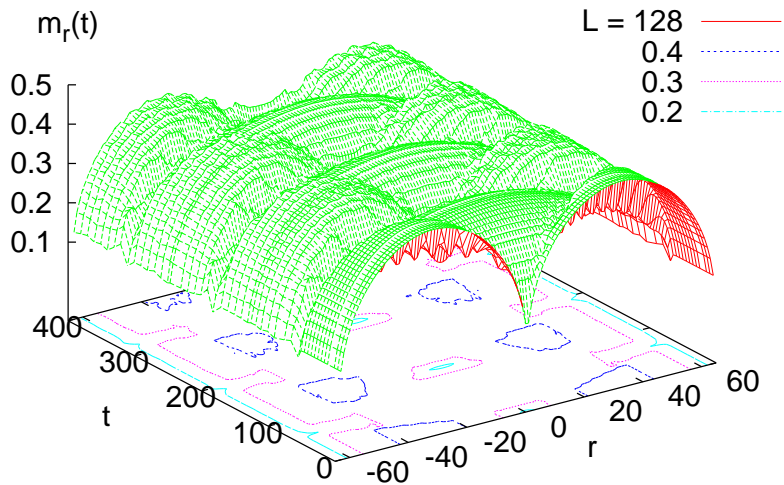}
\caption{
\label{mag3d} (Color online)
Temporal evolution of the local magnetization profiles at the critical point after a local quench with free boundary conditions in the initial state.}
\end{center}
\end{figure}

In this section in the initial state the two half chains have free boundary
conditions, which means that in Eq.(\ref{hamilton}) we have $h_{L/2}=h_{L/2+1}=1$ and $J_{L/2}=0$.
In this settings no conformal results are available, therefore
we study numerically the properties of the profiles of the off-diagonal order parameter, $m_r(t)$ in Eq.(\ref{mag})
in finite systems. The results are shown 
for $L=128$ in Fig. \ref{mag3d}. 

Initially (at $t=0$) the ground state magnetization profiles are
identical and independent in both disconnected parts of the
system. The functional form is known from conformal invariance \cite{ti-97} and
the complete profile has the finite size scaling
form
\be
m_r(t=0,L)\propto L^{-x}\left|\sin\frac{2\pi r}{L}\right|^{x_s-x}\;.
\label{magt0}
\ee
Close to the defect, i.e. for $|r| \ll L/2$ the behavior of the profile
follows from the scaling relation, that $m_r(t=0,L)=b^{-x}m_{r/b}(t=0,L/b)$, with
a rescaling factor, $b>1$. Now taking $b=L$, we arrive to the form:
$m_r(t=0,L)=L^{-x}\widetilde{m}(r/L)$, where the scaling function, $\widetilde{m}(\rho)$ for small argument
behaves as $\widetilde{m}(\rho) \sim \rho^{x_s-x}$. This is compatible with the conformal result
in Eq.(\ref{magt0}), if we replace $\rho$ by its sinusoidal extension: $sin(2\pi \rho)$.

\begin{figure}[t]
\begin{center}
\includegraphics[width=8cm,angle=270]{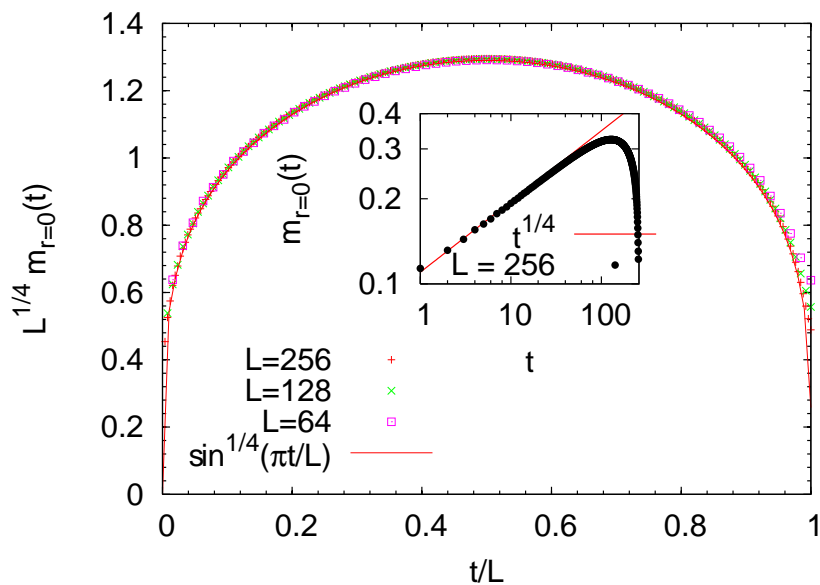}
\caption{
\label{mag-scale-t} (Color online)
Scaling plot of the local magnetization at the
defect, $m_0(t)$, 
after a local quench at criticality for 
different system size $L$. The conjectured result in Eq.(\ref{scale1})
is shown by a full line.
{\bf Inset:} Log-log plot of $m_{0}(t)$ vs. $t$ 
for $L=256$. The straight line is proportional to $t^{1/4}$.
}
\end{center}
\end{figure}

In Fig.\ \ref{mag3d} one observes a quasi-periodic time-dependence of
the profile with the characteristic initial double peak being
exchanged against a single peak at times $T=L/2,3L/2,\ldots$ and
re-occurring at times $T=L,2L,\ldots$. 
Let us focus first on the time-dependence of the magnetization at the central cite,
$m_1(t,L)$, which is expected to satisfy the same type of scaling relation as $m_1^{(+)}(t,L)$,
thus: $m_1(t,L)=b^{-2x}m_1(t/b,L/b)$. As before taking $b=L$ we arrive to:
$m_1(t,L)=L^{-2x}\widetilde{\mu}_1(t/L)$, where the scaling function, $\widetilde{\mu}_1(\tau)$, for small
argument should behave as $\widetilde{\mu}_1(\tau) \sim \tau^{x_s-2x}$. In this way we obtain:
$m_1(t,L)=L^{-x_s}t^{x_s-2x}$, which is in agreement with the $L$-dependence of the magnetization
at the defect at $t=0$, see in Eq.(\ref{magt0}).
Furthermore we have for the relaxation 
for small times: $m_{1}(t)\sim t^{1/4}$,
which agrees well with the numerical data shown in the 
inset of Fig. \ref{mag-scale-t}.
The form of the scaling function can be conjectured using the same substitution, $\rho \to sin(2\pi \rho)$,
as for $t=0$. In this way we obtain:
\be
m_1(t,L)\propto L^{-2x}\left|\sin\frac{2\pi t}{L}\right|^{x_s-2x}
\label{scale1}
\ee
Fig. (\ref{mag-scale-t}) displays a corresponding finite size scaling
plot for the first period which shows a good data collapse (a
corresponding scaling plot for larger values of $t/L$ is equally good,
data not shown) and thus confirms our conjecture (\ref{scale1}), which
probably can be derived rigorously from conformal invariance.

\begin{figure}[t]
\begin{center}
\includegraphics[width=8cm,angle=270]{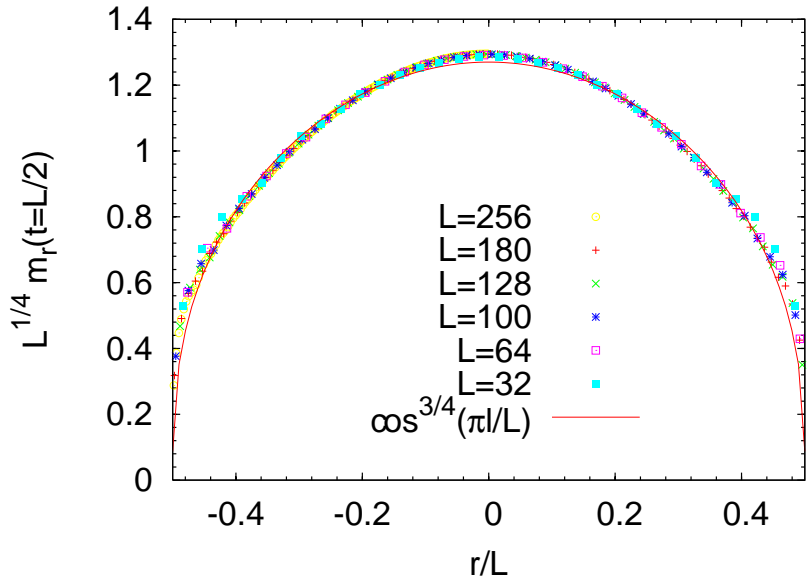}
\caption{
\label{mag-scale-l} (Color online)
Scaling plot of the local magnetization $m_r(t)$ 
after a local quench at criticality for 
different system size $L$ at time $t=L/2$,
see Eq.(\ref{scale2}).}
\end{center}
\end{figure}

Next we consider $t=L/2$, when the profile has its maximum at $r=0$
and it is minimal at the two ends of the chain, $r=\pm L/2$, i.e. at $l=1$ and $l=L$. 
Let us consider the profile for small $l$ and use the scaling transformation:
$m_l(t/L=1/2,L)=b^{-2x} m_{l/b}(t/L=1/2,L/b)$, which leads to the result: $m_l(t/L=1/2,L)=L^{-2x}\mu(l/L)$ , with $b=L$.
The scaling function, $\mu(y)$, for small argument is expected to behave in the same way as for $t=0$, thus
$\mu(y) \sim y^{x_s-x}$. Furthermore having the substitution: $y \to \sin(2\pi y)$ we arrive to the conjecture:
\be
m_{r}(t=L/2,L)\propto L^{-2x}\left|\cos\frac{\pi r}{L}\right|^{x_s-x}\;,
\label{scale2}
\ee
where we use the variable $r$.
Fig. \ref{mag-scale-l} displays a corresponding finite size 
scaling plot which shows a good data collapse and 
thus supports our conjecture (\ref{scale2}).

Combining the scaling forms (\ref{scale1}) and (\ref{scale2})
the ratio $R={m_{l=L/2}^{(L)}(t=\tau)}/{m_{l=\tau}^{(L)}(t=L/2)}$
is given by
\be
R=\left(\sin\pi\frac\tau L\right)^{-x}\;.
\ee

\subsection{Spatial correlations}

The spatial correlation function, $C_t(r_1,r_2)$ in Eq.(\ref{C_t}), has also
been studied by conformal field theory \cite{cc-07loc} and various analytical predictions have
been made in an {\it infinite} system in the continuum limit. In this section we will
compare our results for {\it finite} lattice systems with free
boundaries with these predictions. We note that before the quench
the two halves of the system are disconnected and free, i.e. $h_{L/2}=h_{L/2+1}=1$
and $J_{L/2}=0$. Without the restriction of generality we take
$r_1>|r_2|>0$ and consider two cases: 1) both sites of reference are at the same side of the
defect ($r_2>0$), and 2) the two sites are at different side of the defect ($r_2<0$).

\begin{figure}[t]
\begin{center}
{\bf A}
\includegraphics[width=4.5cm,angle=270]{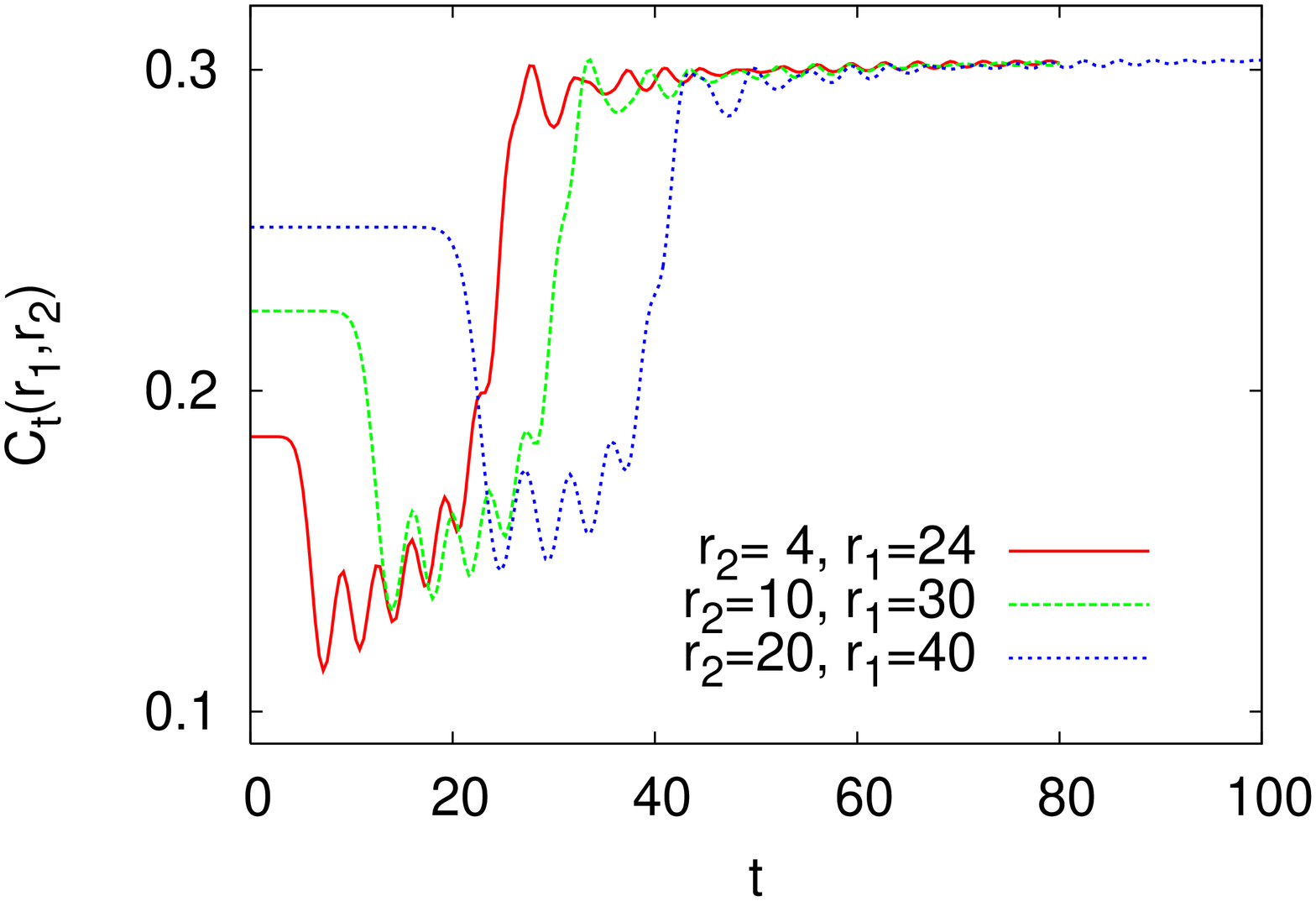}
{\bf B}
\includegraphics[width=4.5cm,angle=270]{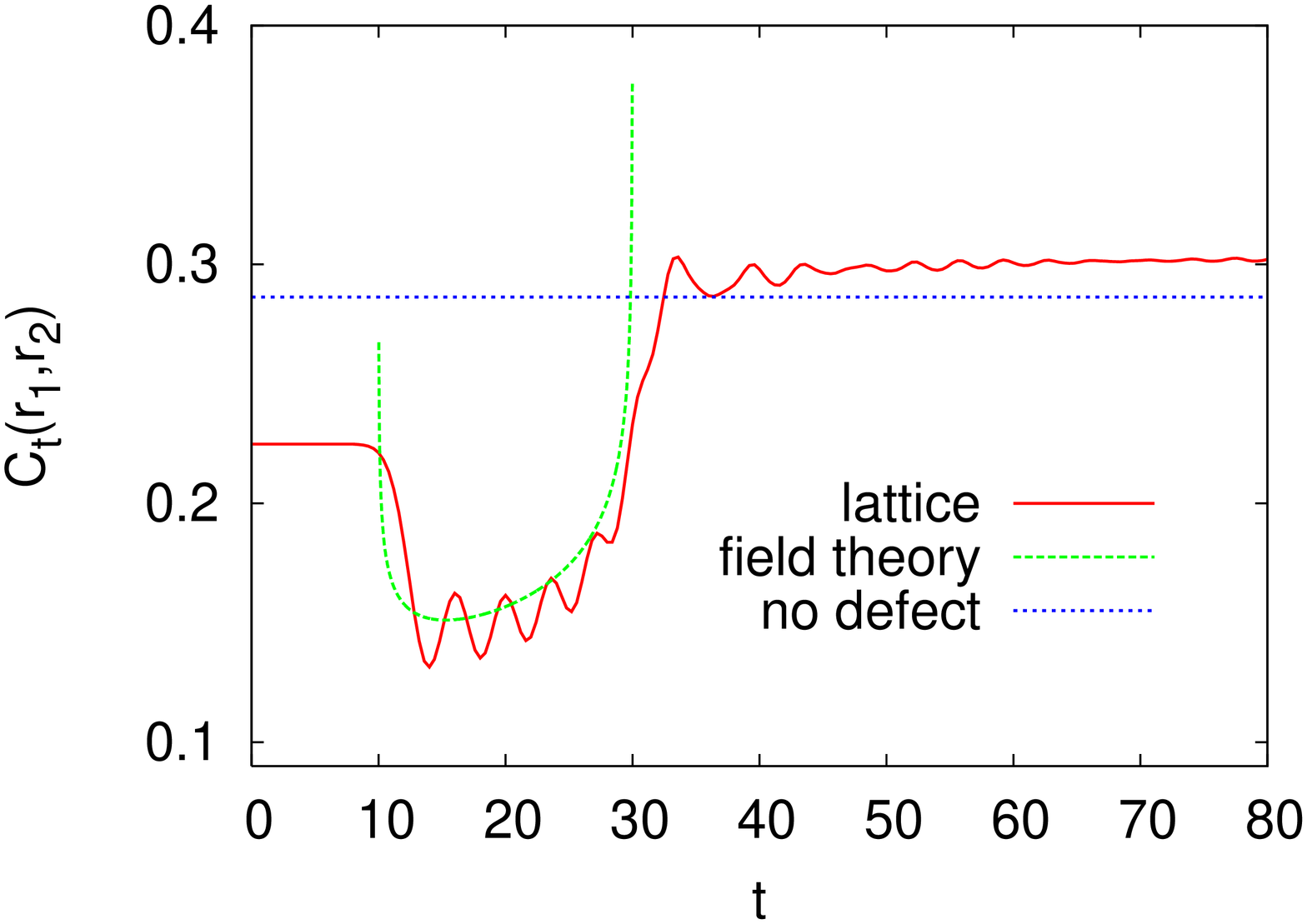}

{\bf C}
\includegraphics[width=4.5cm,angle=270]{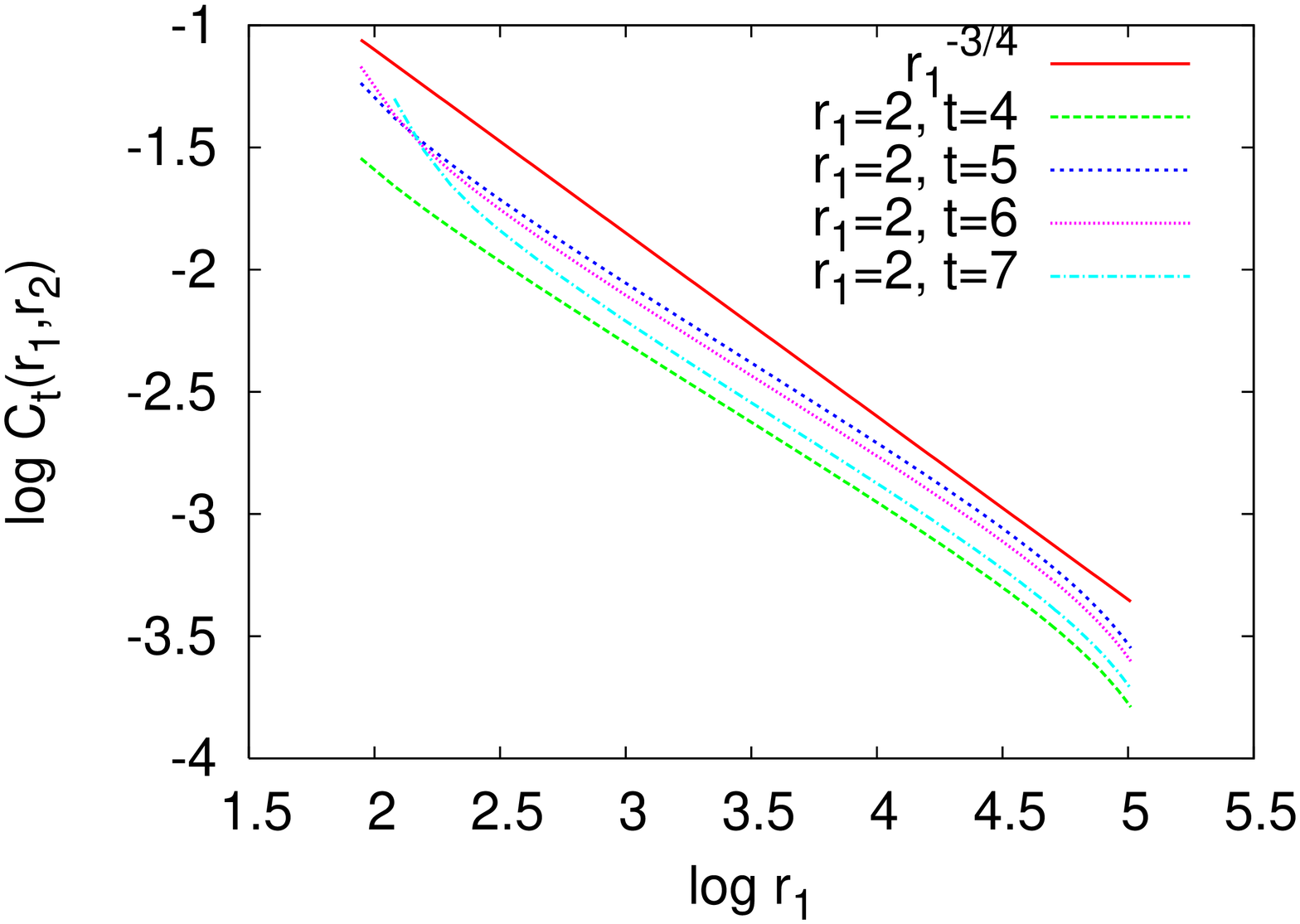}
{\bf D}
\includegraphics[width=4.5cm,angle=270]{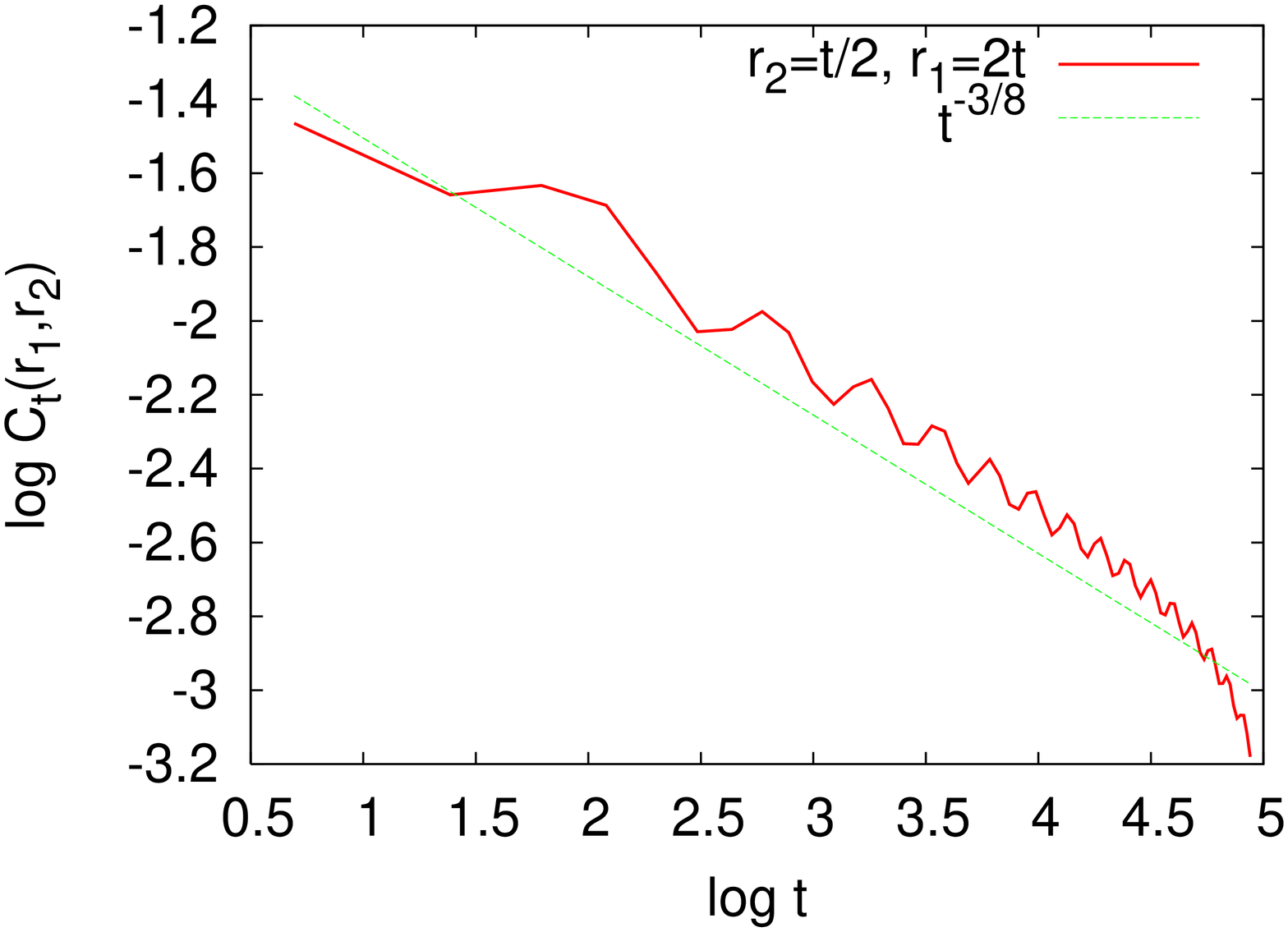}
\caption{
\label{corr-crit-out} (Color online)
{\bf A:} Spatial correlation function 
$C_t(r_1,r_2)$
as a function of time $t$ for both sites of reference being at the
same side of the defect (i.e. $r_1\ge r_2>0$). The system
size is $L=256$, as in the other panels.
{\bf B:} Comparison of the data of A for $r_2=10$, $r_1=30$ with the
conformal field theory result (\ref{corr-out}).
{\bf C:} $C_t(r_1,r_2)$ in the intermediate time regime ($r_2<t<r_1$) for small $r_2$ and $t$
as a function of $r_1$. In the limit $r_1\to\infty$ 
this should approach $r_1^{-3/4}$ according to the
conformal field theory result (\ref{corr-out2}).
{\bf D:} $C(r_1,r_2;t)$ in the intermediate time regime for $r_2=t/2$ and $r_1=2t$ as 
a function of $t$ that should decay as $t^{-3/8}$
according to the conformal field theory result (\ref{corr-out3}).
}
\end{center}
\end{figure}

{\bf Case 1: $r_2>0$} For short times, $t<r_2(<r_1)$, the behavior of the correlation
function can be obtained in the quasi-particle picture. In this case the quasi-particles (which
can be called as signals) starting at the defect
and propagating with a speed $v=1$ does not reach any point of reference, thus the correlations
keep their initial value: $C_t(r_1,r_2)=C_0(r_1,r_2) \propto |r_1-r_2|^{-2x}$. This result follows
also from conformal field theory and consistent with our numerical 
data depicted in Fig. \ref{corr-crit-out} a and b.

For intermediate times, $r_2<t<r_1$, the signals reach the closest site at $r_2$ but not yet 
the remote site at $r_1$, see \ref{corr-crit-out} a and b. The prediction 
from conformal field theory  \cite{cc-07loc} is
\beqn
C_t(r_1,r_2)&=&\left\{
\frac{(r_1+r_2)(r_2+t)\epsilon}{(r_1-r_2)(r_1-t)4r_1(t^2-r_2^2)}
\right\}^x  \nonumber \\
&\times& F\left(\frac{2r_1(r_2+t)}{(r_1+r_2)(r_1+t)}\right)  
\label{corr-out}
\eeqn
with 
\be
F(\eta)=\left(\sqrt{1+\sqrt{\eta}}-\sqrt{1-\sqrt{\eta}}
\right)/\sqrt{2}
\sim\sqrt{\eta}\quad{\rm for}\;\eta\to0
\label{FF}
\ee 
and $\epsilon$ is the regularization parameter in Eq.(\ref{Joukowsky}).
In Fig.\ \ref{corr-crit-out}b a comparison with
our numerical data for a specific $r_1$ and $r_2$ is shown.
As expected for a lattice model our data display 
characteristic oscillations in the considered regime
around the monotonous continuum result (\ref{corr-out}).

For a more quantitative comparison we first look at 
the asymptotic behavior of $C_t(r_1,r_2)$ for fixed 
small $t$ and $r_2$. The conformal field theory 
predicts according to (\ref{corr-out}) and (\ref{FF})
\be
C_t(r_1,r_2)\propto r_1^{-2x-x_s}=r_1^{-3/4}\quad
{\rm for}\;t<r_2\ll r_1
\label{corr-out2}
\ee
which agrees well with our lattice result as shown 
in Fig.\ \ref{corr-crit-out}c. Moreover, for fixing
the ratios $\rho_2=r_2/t<1$ and $\rho_1=r_1/t>1$
(\ref{corr-out}-\ref{FF})
predicts in the limit $t\to\infty$
\be
C_t(r_1,r_2)\propto t^{-3x}=t^{-3/8}\quad
{\rm for}\; \rho_2<1\; \rho_1>1\;t\to\infty\;.
\label{corr-out3}
\ee
This is also consistent with our finite lattice data as seen
in Fig.\ \ref{corr-crit-out}d for $\rho_2=1/2$ and $\rho_1=2$.

In the long-time regime, for $t>r_1$ the signals have also reached the remote site at
$r_1$, and both points of reference has the information of the joining of
the two halves of the system. Consequently
$C_t(r_1,r_2)$ approaches a time-independent value
(c.f.\ Fig.\ \ref{corr-out}a and b), given by the equilibrium bulk correlation function,
which is proportional to $|r_1-r_2|^{-2x}$.

\begin{figure}[t]
\begin{center}
{\bf A}
\includegraphics[width=4.5cm,angle=270]{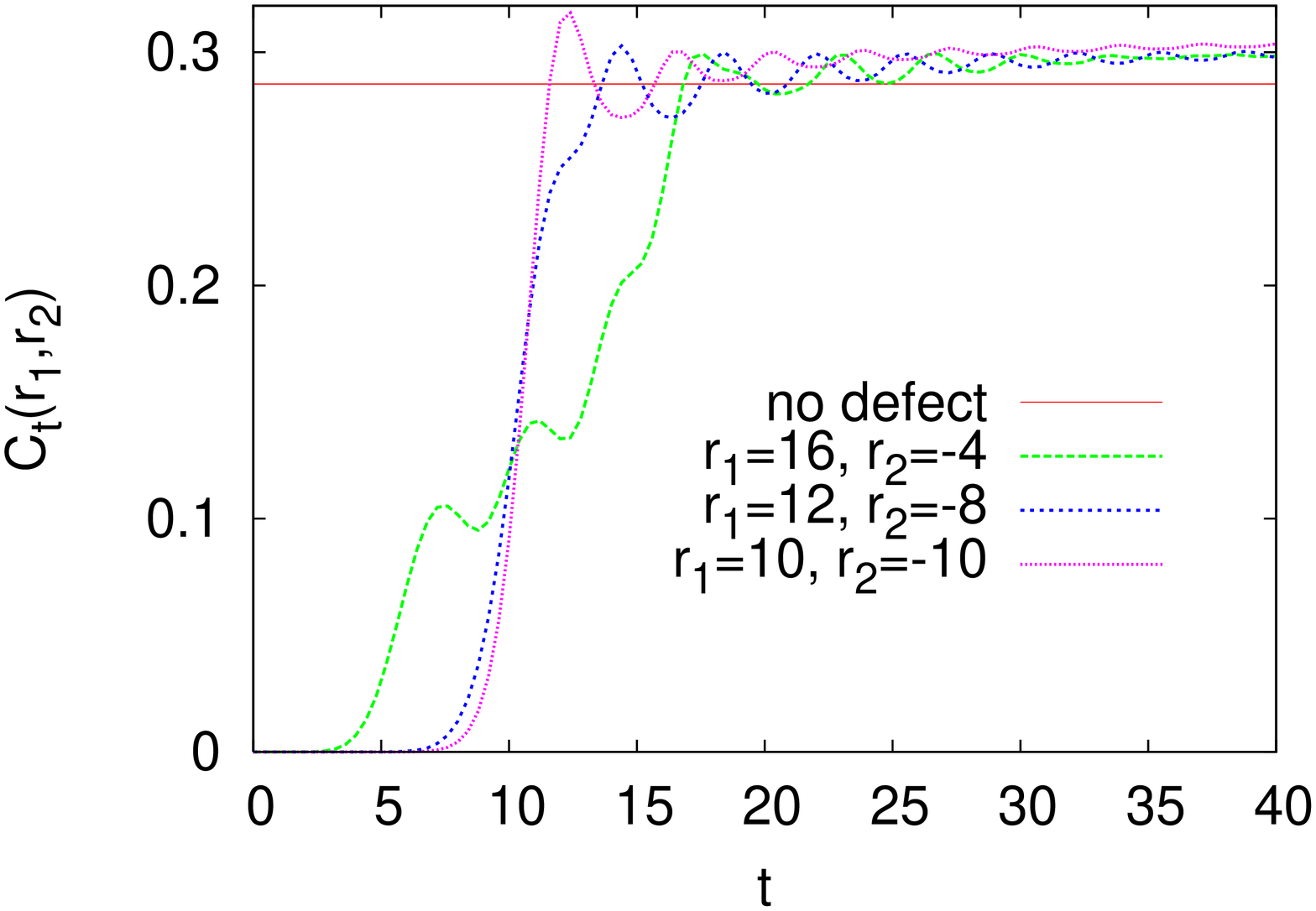}
{\bf B}
\includegraphics[width=4.5cm,angle=270]{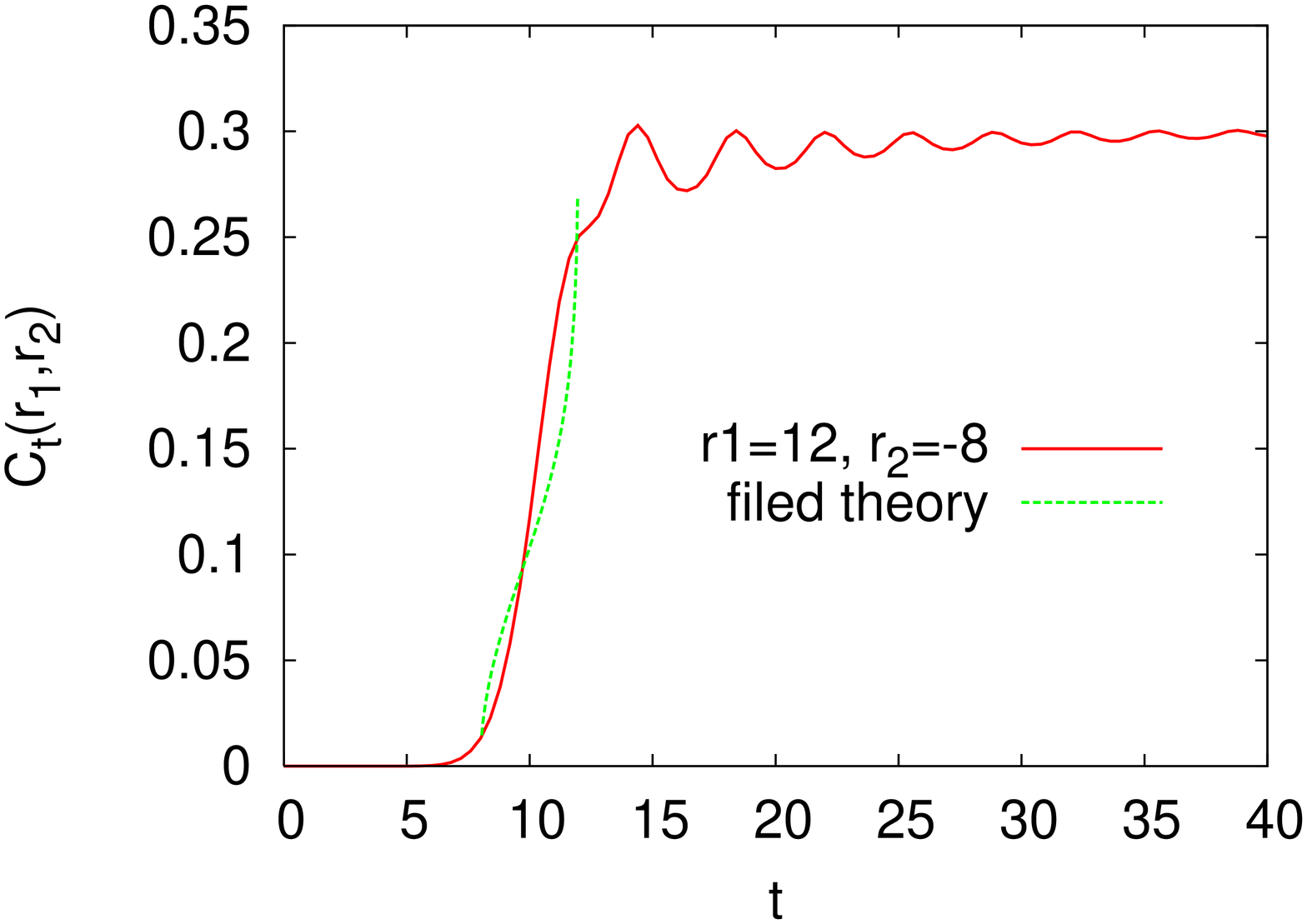}

{\bf C}
\includegraphics[width=4.5cm,angle=270]{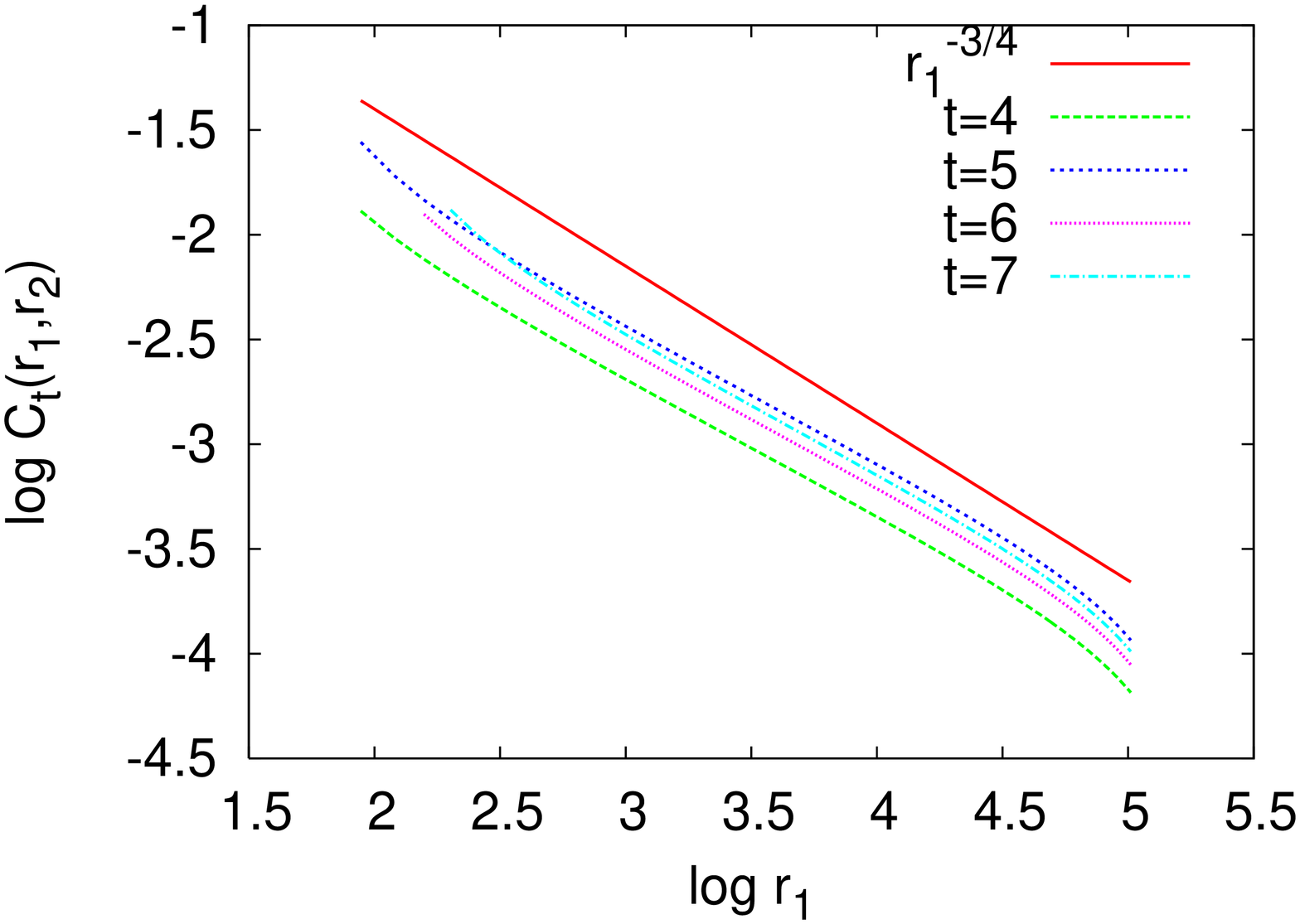}
{\bf D}
\includegraphics[width=4.5cm,angle=270]{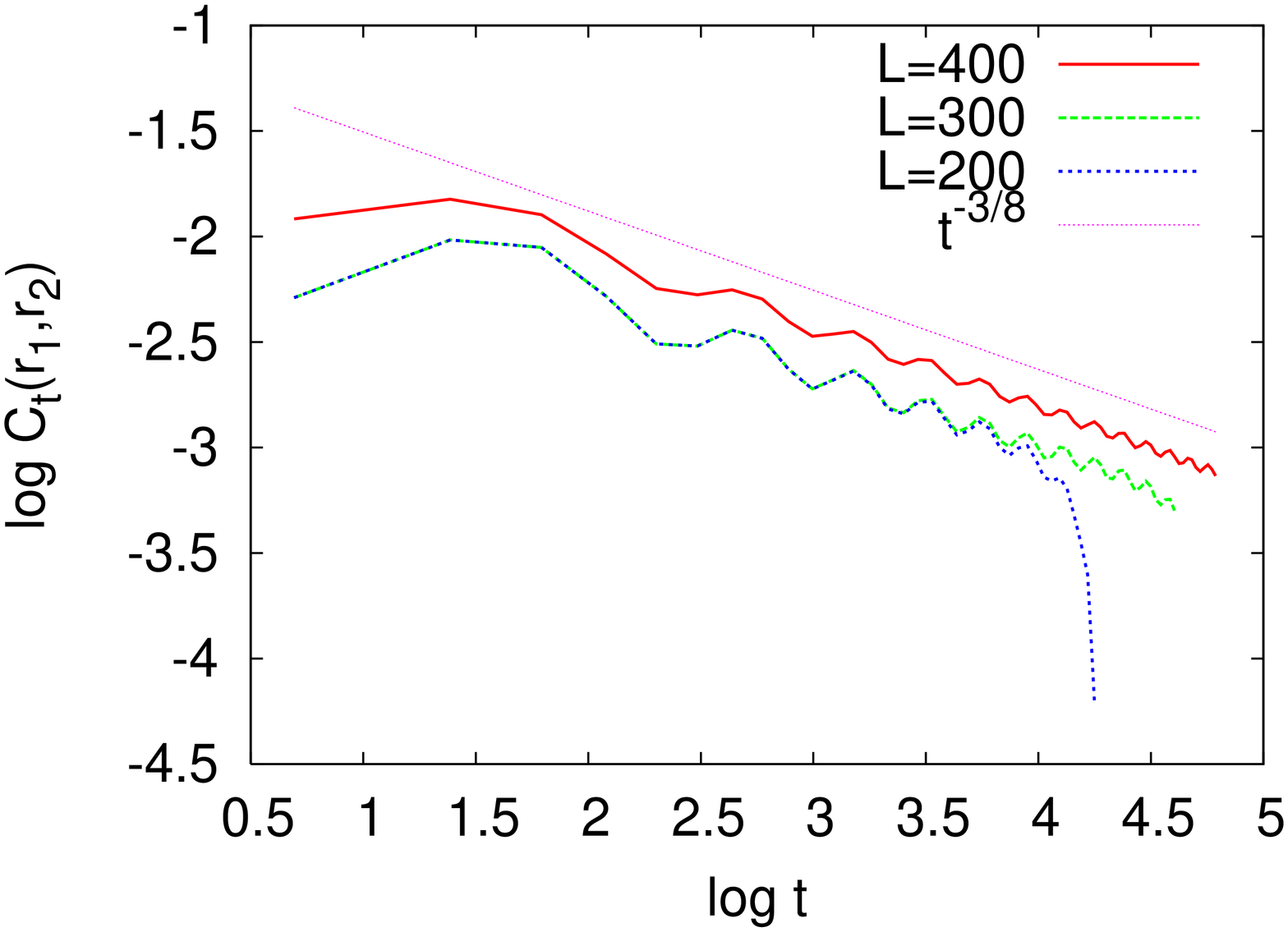}

\caption{
\label{corr-crit-in} (Color online)
{\bf A:} Spatial correlation function 
$C_t(r_1,r_2)$
as a function of time $t$ for the two sites of reference being at different
sides of the defect: $r_1>0$, $r_2<0$ ($r_1\ge |r_2|$). The system
size is $L=256$, as in the other panels.
{\bf B:} Comparison of $C_t(r_1,r_2)$ with the
conformal field theory result (\ref{corr-in}).
{\bf C:} $C_t(r_1,r_2)$ for small $|r_2|$ and $t$
as a function of $r_1$. In the limit $r_1\to\infty$ 
this should approach $r_1^{-3/4}$ according to the
conformal field theory result (\ref{corr-out2}).
{\bf D:} $C_t(r_1,r_2)$ for $r_2=t/2$ and $r_1=2t$ as 
a function of $t$ that should decay as $t^{-3/8}$
according to the conformal field theory result (\ref{corr-out3}).
}
\end{center}
\end{figure}

{\bf Case 2: $r_2<0$}, i.e. the two reference sites are at the different sides
of the defect. Initially, at $t=0$, the
two magnetization operators are located in separate, independent subsystems, thus their
correlations should vanish. As long as after the quench no
signal reaches the site at $r_2$ (i.e.\ for $t<|r_2|$) the
correlation function should be expected to be zero. (Strictly speaking
$C_t(r_1,r_2)$ can be exponentially small, the only
contributions coming from signals propagating outside the
``light-cone''.) This is clearly visible in our results shown in 
Fig.\ \ref{corr-crit-in}a. Surprisingly this disagrees with the 
conformal field theory result  \cite{cc-07loc} which 
predicts
\be
C_t(r_1,r_2)=\frac{1}{|4r_1r_2|^x}\cdot
F\left(\frac{\epsilon^2r_1|r_2|}{(r_1^2-t^2)(r_2^2-t^2)}
\right)\quad{\rm for}\;t<|r_2|\;.
\ee
This could be rectified by setting $\epsilon=0$, but this 
would be at variance with the results for $r_2>0$, for
which the regularization parameter should be non-vanishing. The
problem with the conformal derivation could be related to the fact,
that in the transformation (\ref{Joukowsky}) points in the $z$ plane
with ${\rm Re}(z_2)=r_2<0$ and $|r_2| \gg \epsilon$ are transformed
to $|w_2| \ll \epsilon$. However, the scaling form of the correlation
function in the semi-infinite $w$-plane is valid only in the
continuum limit, i.e. for  ${\rm Re}(w_2) \gg 1$.

After the signals reach the site at $r_2$, i.e.\ for $|r_2|<t<r_1$,
the conformal field theory prediction is  \cite{cc-07loc}
\beqn
&C_t(r_1,r_2)=\left(
\frac{(r_1+r_2)(r_2+t)\epsilon}{(r_1-r_2)(r_1-t)4r_1(t^2-r_2^2)}
\right)^x \nonumber \\
&\times
F\left(\frac{2r_1(r_2+t)}{(r_1+r_2)(r_1+t)}\right)
\quad{\rm for}\;
|r_2|<t<r_1\;.
\label{corr-in}
\eeqn
In Fig.\ \ref{corr-crit-in}b a comparison with
our numerical data for a specific $r_1$ and $r_2$ is shown.
As expected for a lattice model our data display 
characteristic oscillations in the considered regime
around the monotonous continuum result (\ref{corr-in}).

For fixed small $t$ and $r_2$ the conformal field theory predicts
according to (\ref{corr-in}) and (\ref{FF}) again for $C_t(r_1,r_2)$
the asymptotic $r_1$-behavior (\ref{corr-out2}) as in the case
$r_2>0$. This agrees well with our lattice result as shown in Fig.\
\ref{corr-crit-in}c. Moreover, for fixed $\rho_2=r_2/t>-1$ and $\rho_1=r_1/t>1$
(\ref{corr-in}) predicts in the limit $t\to\infty$ the same asymptotic
$t$-behavior (\ref{corr-out3}) as in the case $r_2>0$. This is also in
agreement with our  lattice result as shown in Fig.\
\ref{corr-crit-in}d.
Finally for $t>r_1$ the signal has also reached the site at $r_1$
and $C_t(r_1,r_2)$ approaches again the time-independent equilibrium bulk value
(c.f.\ Fig.\ \ref{corr-crit-in}), which is proportional to $|r_1-r_2|^{-2x}$.

Summarizing our results confirm the predictions of the conformal field theory 
 \cite{cc-07loc} for the spatial correlation after a 
local quench at the critical point, except for the case $r_2<0$,
$t<|r_2|$.

\subsection{Autocorrelations}

We have also calculated the autocorrelation function, $G_r(t,0)$, as defined in Eq.(\ref{G_r}).
In the following for simplicity we shall omit the second argument and use the notation $G_r(t)$.
Using the quasi-particle picture we have the following expectations. Before the signal reaches the reference point,
$t<r$, the autocorrelation function is the same as in the equilibrium bulk system, thus asymptotically
$G_r(t) \propto t^{-2x/z}=t^{-1/4}$. For $t>r$, when the signal has passed the reference point the equilibrium
bulk decay of $G_r(t)$ should continue, thus in the complete time-window this behavior is observable.

\begin{figure}[h!]
\begin{center}
\includegraphics[width=6cm,angle=270]{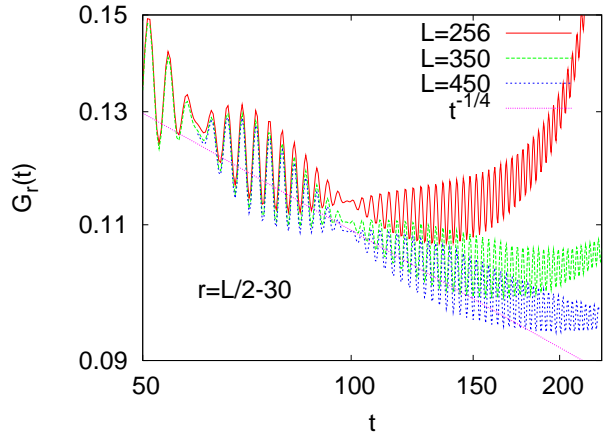}
\caption{
\label{auto-crit} (Color online)
Autocorrelation function 
$G_r(t)$ after a critical quench for different system sizes.
The straight line corresponds to the equilibrium power law
$t^{-1/4}$.
}
\end{center}
\end{figure}

Our lattice results in Fig.\ \ref{auto-crit} are in agreement with these expectations. In a finite lattice
of length $L$ the autocorrelation function has a minimal value of $\sim L^{-1/4}$.

\section{Quench in the ferromagnetic phase}
\label{sec:ferro}
\begin{figure}[h!]
\begin{center}
\includegraphics[width=12cm]{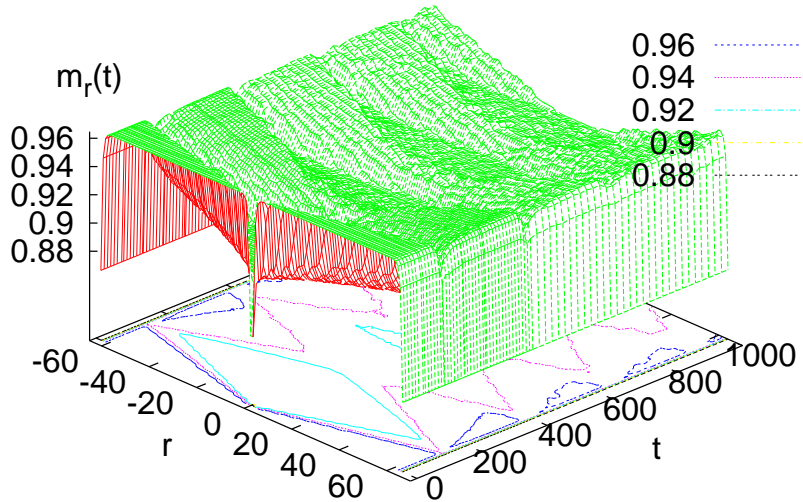}
\caption{
\label{mag-FM-3d} (Color online)
Temporal evolution of the local magnetization profiles after 
a local quench within the ordered (FM) phase, here $h=0.5$, $L=128$.
}
\end{center}
\end{figure}

In equilibrium in the ferromagnetic (FM) phase ($h<1$) there is spontaneous
order and the bulk $m_b$ and the surface, $m_s$, magnetizations are
given by:
\be
m_b=(1-h^2)^{x \nu},\quad m_s=(1-h^2)^{x_s \nu},
\ee
respectively (the correlation length exponent is $\nu=1$). The magnetization profile has an exponential
variation in the surface region, the size of which is given
by: $\xi_s \sim (1-h)^{-1}$, close to the critical point. Here we follow the
same protocol as in Sec.\ref{sec:magn_free}: for $t<0$ we cut the system
into two halves with free boundary conditions, which are then (at $t=0$) joined together
with the coupling $J_{L/2}=J=1$. We measure the temporal evolution of the magnetization profile, $m_r(t)$,
as defined in Eq.(\ref{mag}). The finite lattice results are depicted
in Fig.\ \ref{mag-FM-3d} for $h=0.5$ and $L=128$. In the initial state
the magnetization profile ($m_r(t=0)$)
is essentially constant and given by $m_b$, except close to the boundaries and to the center.
For $t>0$ one observes again a quasi-periodic pattern and the period of time $T(h)$ is found
to increase with increasing $L$ and decreasing 
field $h$. It turns out that the spatio-temporal evolution of the
profile can be understood even quantitatively within a 
quasi-classical picture.

As elaborated in  \cite{igloi_rieger3} within the FM phase the dynamics
of the local order parameter in a system with boundaries after a {\it
global} quench is very well described by a quasi-particle (QP) picture, 
where QPs are kinks that move with velocity $\pm v_p$ through the
system. The QP energy is
\be
\epsilon_p=\sqrt{1+h^2-2h\cos(p)}
\ee
with $p=(2 n-1)\pi/2L$ ($n=1,\ldots,L$) and the QP velocity is
\be
v_p=\frac{\partial\epsilon_p}{\partial p}
=\frac{h\sin(p)}{\epsilon_p}\;.
\ee
The maximum velocity is given by:
\be
v_{\rm max}^2=\left[\sqrt{(1+h^2)^2+12h^2} - (1+h^2)\right] /6\;,
\ee
for small $h$ it is $v_{\rm max} \approx h$.

Consider now a QP, or kink, with momentum $p$. It moves uniformly 
with velocity $v_p$ until it reaches one of the boundaries, where it
is reflected and moves with velocity $-v_p$ thereafter, and so forth.
The trajectory of the kink is periodic in time, after a time $2T_p$
with 
\be
T_p=L/v_p
\ee
(including a reflection at the right and left boundary) it returns 
to the starting point $x_0$ with the initial direction and velocity 
$v_p$. After a {\it global} quench QPs emerge pairwise at
random positions $x_0$ with velocities $+v_p$ and $-v_p$
 \cite{rossini,igloi_rieger3} and therefore we assume that 
after the {\it local} quench studied here QPs emerge also pairwise,
but exclusively at the central site, where the defect is located
before the quench.

The time dependent decay of the local magnetization $m_r(t)$ at a position
$r$ is then determined by the probability $q$ with which any given
kink trajectory passes until time $t$ the site $r$ an odd
(!) number of times: Each passage of one of the two trajectories flips
the spin at site $r$ and an even number of flips does not change the
magnetization of site $r$. Once $q(t)$ is known the magnetization is
given by
\be
m_r(t)=m_r(t=0)\cdot\exp(-2q(t))\;.
\label{mt}
\ee
The probability $q(t)$ is expressed as
\be
q(t)=\frac1\pi \int_0^\pi dp\, f_p(h)\,q_p(t)
\label{qt}
\ee
where $q_p(t)$ is the probability that any given QP trajectory passes
the site $r$ an odd number of times, and $f_p(h)$ is the probability with
which QPs with momentum $p$ are generated (per site), and we take
assume that it is identical to $f_p$ after a global quench. This is given
for small $h$ as \cite{igloi_rieger3}:
\be
f_p(h)=\frac14 h^2\sin^2(p)
\label{fp}
\ee

To calculate $q_p$ one concentrates first on times $t<2T_p$
and on lattice site $r<0$ - the whole profile is of course symmetric
with respect to a reflection at the center $m_{r}(t)=m_{-r}(t)$. The
QPs emerging at the central site $r=1$ and moving to the left
($v_p<0$) pass the site $r$ two times (once before the reflection at
the left boundary and once after the reflection) before they return
(now with $v_p>0$) to their starting point. These two times are
$t_1=|r|/v_p$ and $t_2=T_p-t_1$. Therefore the probability that
this QP trajectory passes $r$ at times $t<T_p$ exactly once is
\be
q_p^-=\left\{
\begin{array}{cl}
0 & t<t_1\\
1 & t_1<t<t_2\\
0 & t_2<t<T_p
\end{array}
\right.
\label{qp1}
\ee
The associated partner (of the QP pair) that moves to the right
($v_p>0$) passes the site $r$ only after reflection at the right
boundary and returns to the starting point, at times $t_3=T_p+t_1$
and $t_4=2T_p-t_1$. Therefore the probability that
this QP trajectory passes $r$ at times $t<2T_p$ exactly once is
\be
q_p^+=\left\{
\begin{array}{cl}
0 & t<t_3\\
1 & t_3<t<t_4\\
0 & t_4<t<2T_p
\end{array}
\right.
\label{qp2}
\ee
For $t>2T_p$ one makes use of the
$T_p$-periodicity of $q_p(t)$: 
\be
q_p(t+2nT_p)=q_p(t),\quad(n=1,2,\ldots).
\label{qptn}
\ee
With (\ref{qt}), (\ref{fp}), (\ref{qp1}-\ref{qptn}) one obtains
$m_l(t)$ via numerical integration (or summation over the discrete
$p$-values for a lattice of finite size $L$), where we take the exact
profile before the quench ($m_l(t=0)$). 

\begin{figure}[t]
\begin{center}
{\bf A}
\includegraphics[width=7cm]{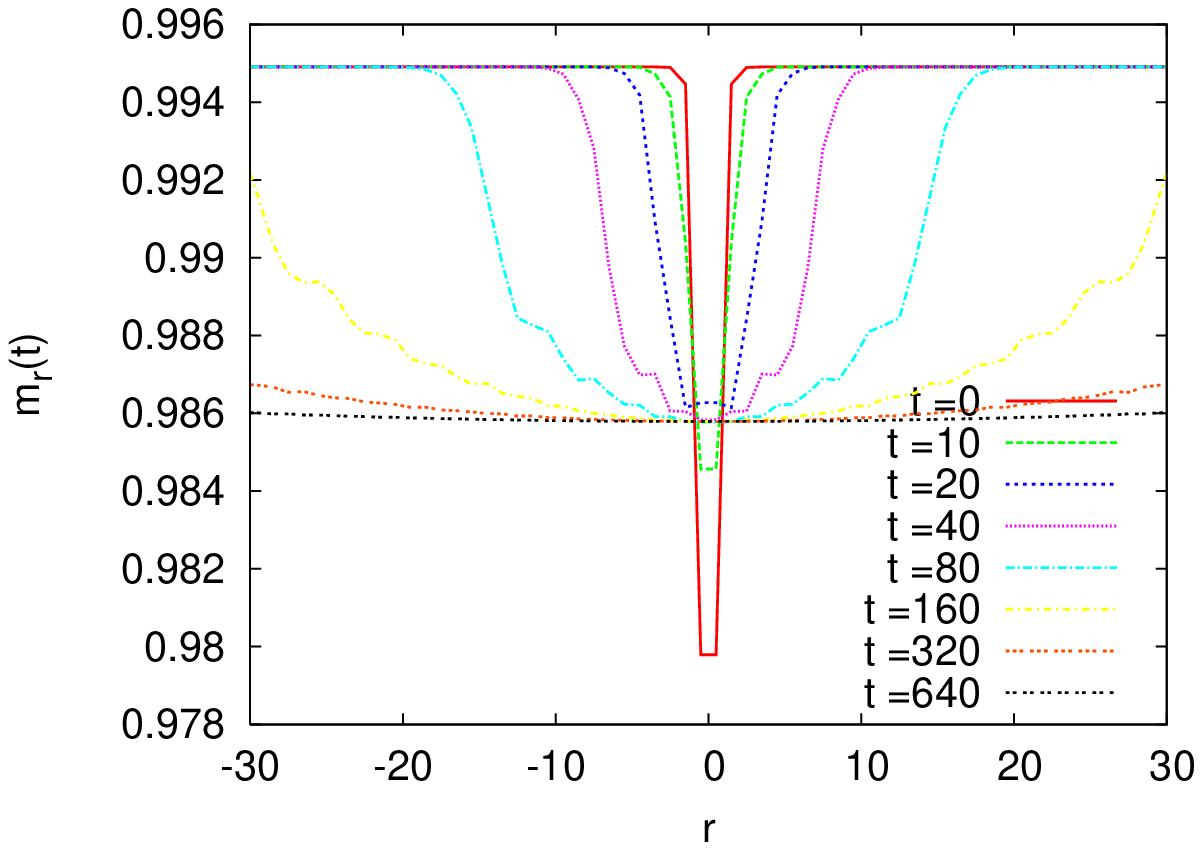}
{\bf B}
\includegraphics[width=7cm]{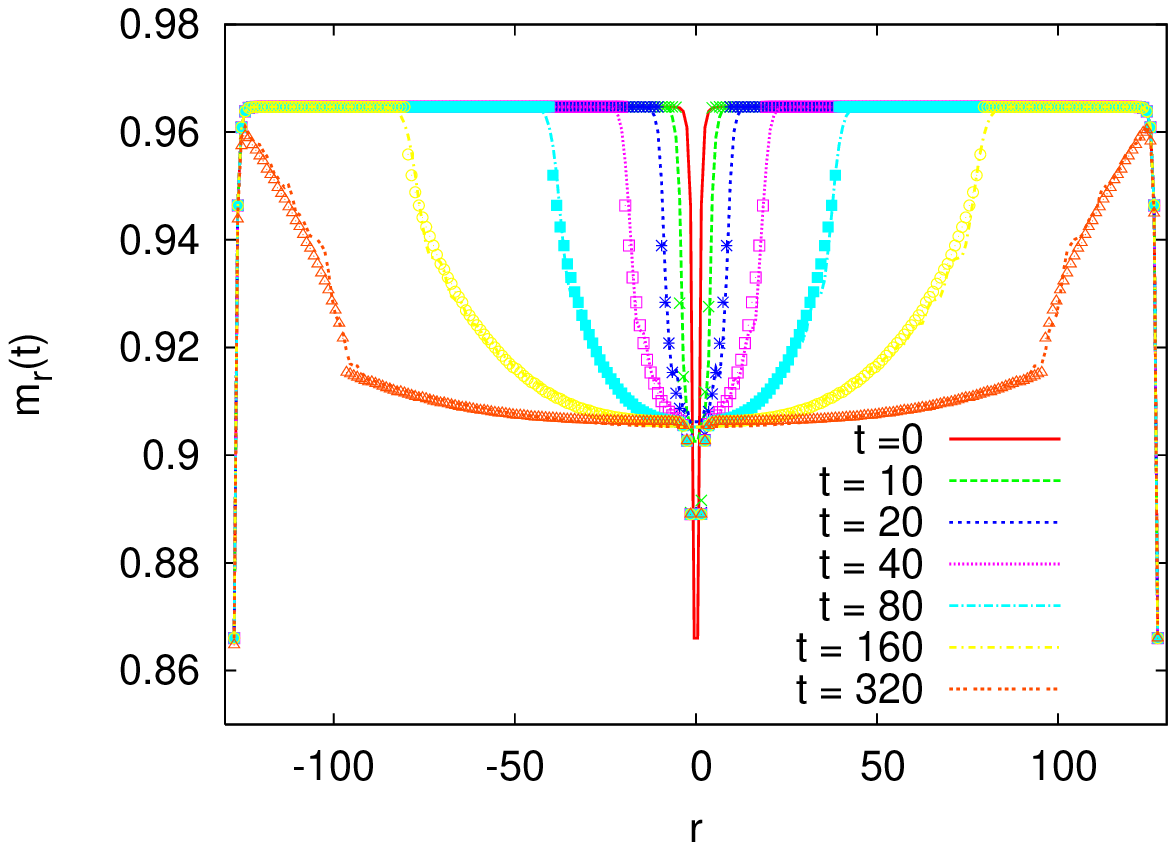}

{\bf C}
\includegraphics[width=7cm]{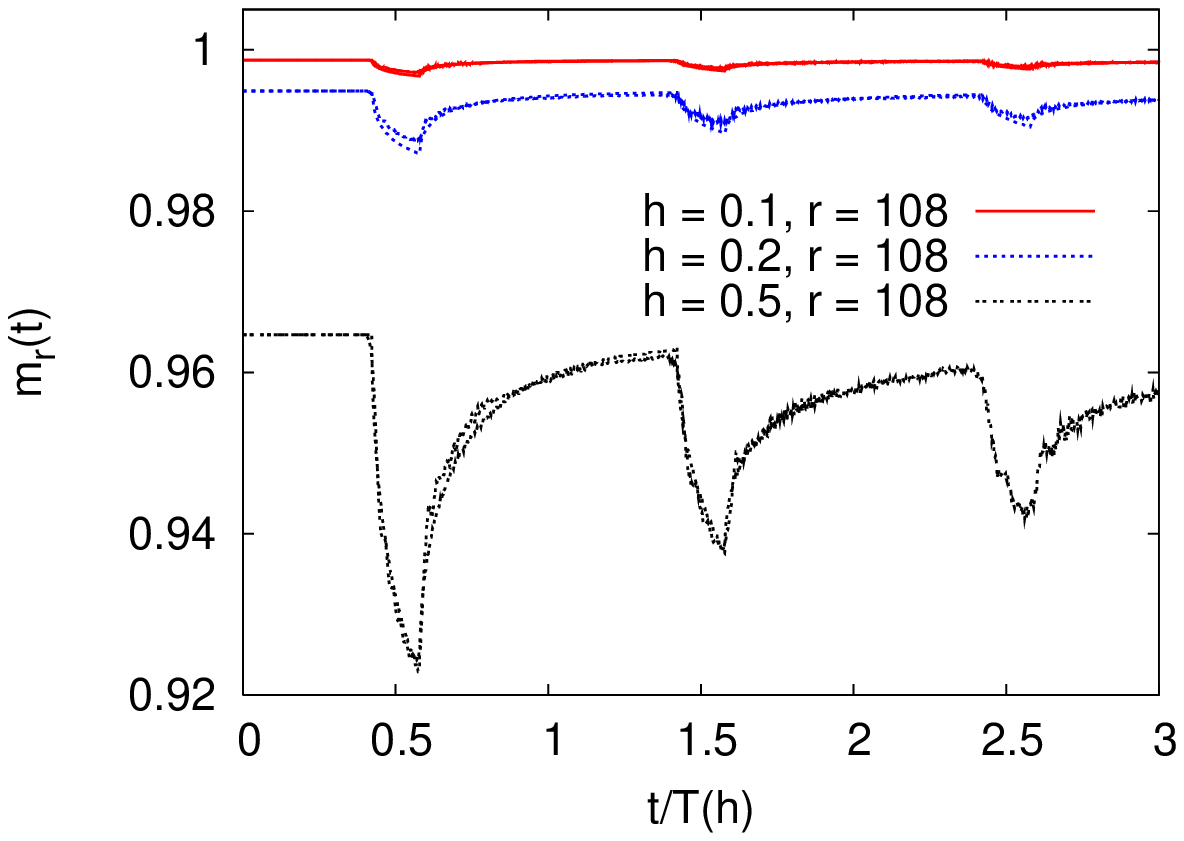}
{\bf D}
\includegraphics[width=7cm]{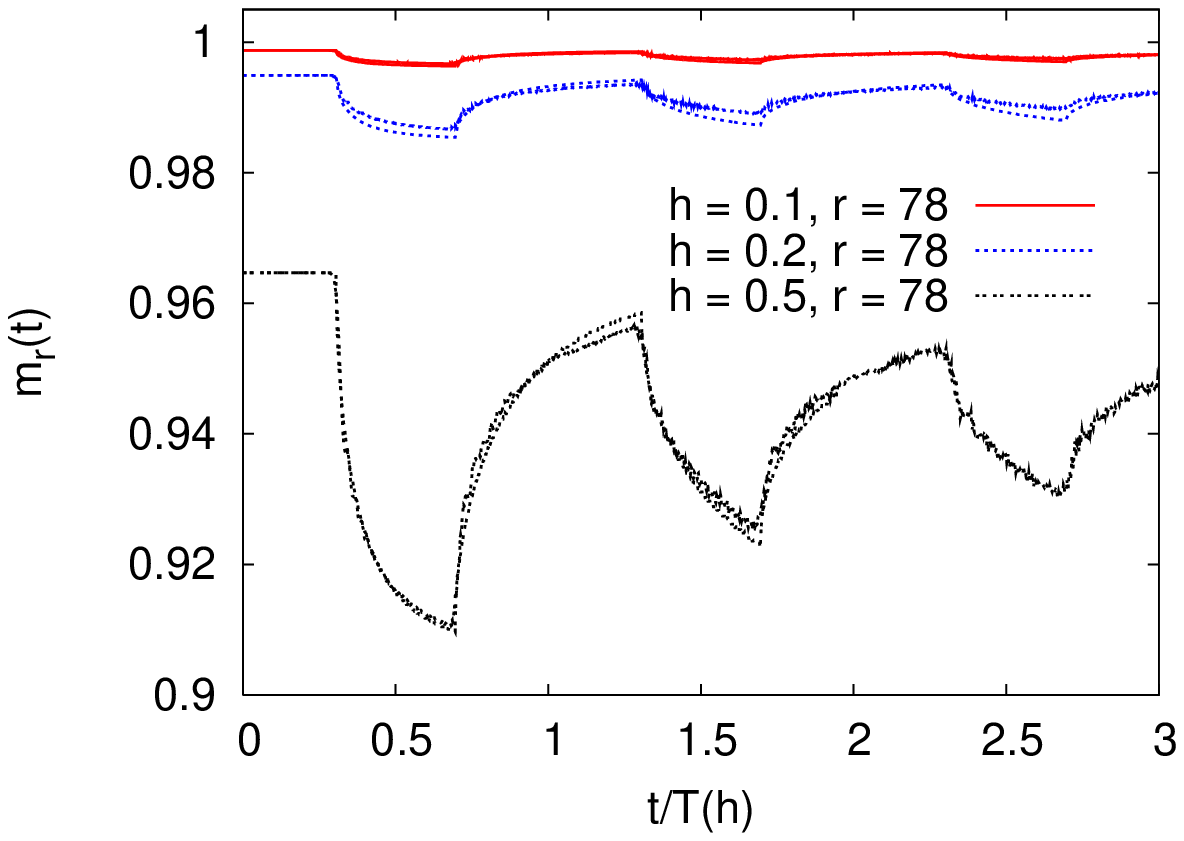}
\caption{
\label{mag-FM} (Color online)
Local order parameter $m_l(t)$ within the FM phase ($h<1$) and
comparison with QP calculation.  {\bf A:} $m_r(t)$ at $h=0.2$ as a
function of the distance $r$ from the defect for different fixed times
$t$. {\bf B:} $m_l(t)$ at $h=0.5$ as a function of $l$ for different
fixed times ($L=256$, the defect is at $l=128$). The full lines are
the exact data and the points are the predictions of the QP
calculation.  {\bf C-D:} $m_l(t)$ for different fields $h<1$ as a
function of the rescaled time $t/T(h)$, where $T(h)=L/v_{\rm
max}(h)=L/h$, i.e.\ $T(h=0.5)=512$, $T(h=0.2)=2560$, and
$T(h=0.1)=5120$. The exact data and the QP prediction can be
discriminated by the smoother behavior of the latter.  }
\end{center}
\end{figure}

In Fig.\ \ref{mag-FM} the comparison of this QP calculation with our
exact data is shown.  Fig.\ \ref{mag-FM}A displays the dynamical
evolution of the local magnetization at $h=0.2$ close to the
defect. The system size, $L=256$, is sufficiently large such that the
boundary effects are not yet visible for the times shown and the data
are representative for an infinite system. One sees that the initial
magnetization drop at the defect (the defect spins are surface spins
at $t=0$) is quickly filled and the profile approaches a constant
magnetization $\widetilde{m}_b=\lim_{L,t\to\infty} m_{r}(t)$.  Since
the probability $q_p(t)$ for a QP with momentum $p$ to pass any site in
the bulk is $1$ in the the limit $L\to\infty$ and $t\to\infty$ 
the bulk magnetization is predicted by the QP picture, according 
to (\ref{qt}) and (\ref{mt}), to be
\be
\widetilde{m}_b=m_b \exp(-1/\xi)\;,
\label{tilde_m_b}
\ee
with
\be
\dfrac{1}{\xi}=\frac1\pi \int_0^\pi dp f_p(h)\;.
\ee
$\xi$ is identical with the nonequilibrium correlation length
measured during \textit{global quench}, as defined in
Eq.(\ref{xi}). For $f_p(h)$ as in Eq.(\ref{fp}), which is exact for
small $h$ it is given by: $1/\xi=h^2/4$. For larger values of $h$
corrections to the kink-like excitations should by summed, which leads
to the value \cite{sps-04,cef-11}:
$1/\xi=-\ln\left[\left(1+\sqrt{1-h^2}\right)/2 \right]$. For $h=0.2$
one gets $\widetilde{m}_b=0.985$ which agrees well with the
approximately constant magnetization at $t=640$ in Fig.\
\ref{mag-FM}A. We checked that also for larger values of $h$ the QP
prediction for the asymptotic bulk magnetization $\widetilde{m}_b$ in
eq.(\ref{tilde_m_b}) is good.

Fig.\ \ref{mag-FM}B compares the exact data for the local
magnetization profile of a finite system for different times after the
quench with those of the QP prediction at $h=0.5$. The agreement is
very good for the times shown and a few sites away from the defect
($r=|L/2-l|>5$). The rapid filling of the initial sharp dip of the 
profile at the center (see Fig.\ \ref{mag-FM}A) is
not correctly captured by the present QP picture. To understand this
process one should assume, that the QP-s are emitted in a region of size $\xi_s$
around the defect and these QP-s are quantum entangled and these
correlated particles are responsible
for the rapid increase of the magnetization at the defect.

The predicted (quasi)-period is governed by the maximum QP velocity
and given by
\be
T(h)=L/v_{\rm max}\approx L/h\;.
\ee
This agrees well with the dynamical behavior of $m_l(t)$ shown in 
\ref{mag-FM} C-D displaying the exact data and the QP prediction for 
$m_l(t)$ at fixed sites $l$ for different fields $h<1$ as a function
of the rescaled time $t/T(h)$. The agreement for the sites shown is
good, at longer times close to the defect ($l\approx L$) deviations
occur (see Fig.\ \ref{mag-FM}D) due to the mechanism described above.

\section{Quench in the paramagnetic phase}
\label{sec:para}

\begin{figure}[h!]
\begin{center}
\includegraphics[width=12cm]{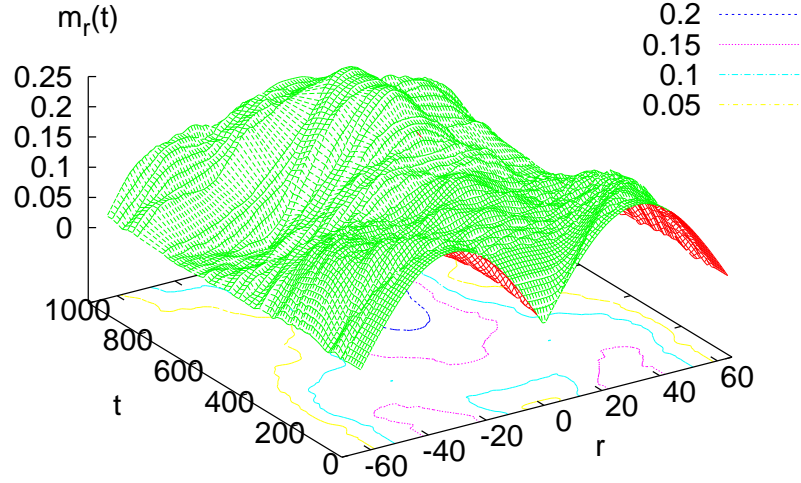}
\caption{
\label{mag-PM-3d} (Color online)
Temporal evolution of the local magnetization profiles after 
a local quench within the disordered (PM) phase, here $h=2$, $L=128$.
}
\end{center}
\end{figure}

In the paramagnetic phase ($h>1$) the magnetization is vanishing in
the thermodynamic limit, however, in a finite system there is a
finite, $L$-dependent magnetization. Its temporal evolution for $h=2$
is depicted in Fig.\ \ref{mag-PM-3d} for $L=128$.  In contrast to the
dynamics at the critical point (c.f.\ Fig.\ \ref{mag3d}) the evolution
of $m_l(t)$ is not periodic in the paramagnetic phase and approaches a
stationary profile characteristic for the system without defect.

\begin{figure}[h!]
\begin{center}
{\bf A}
\includegraphics[width=6cm]{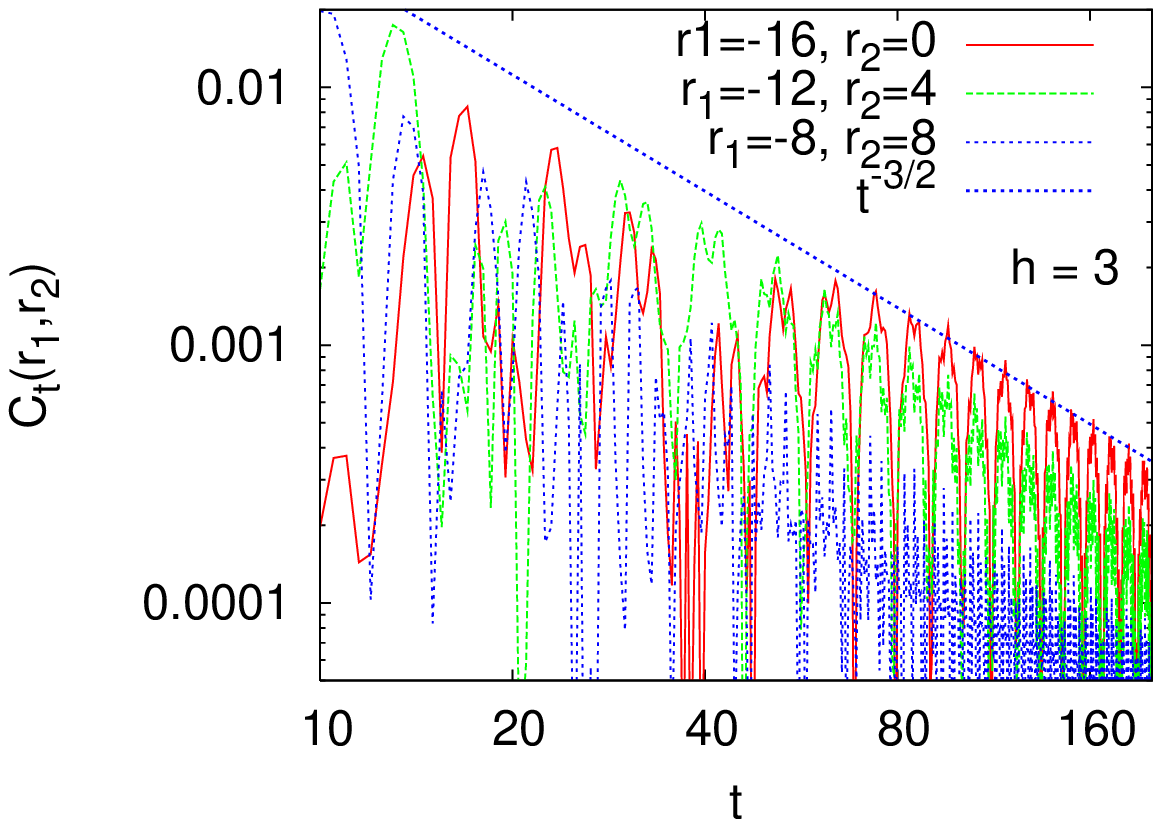}
{\bf B}
\includegraphics[width=6cm]{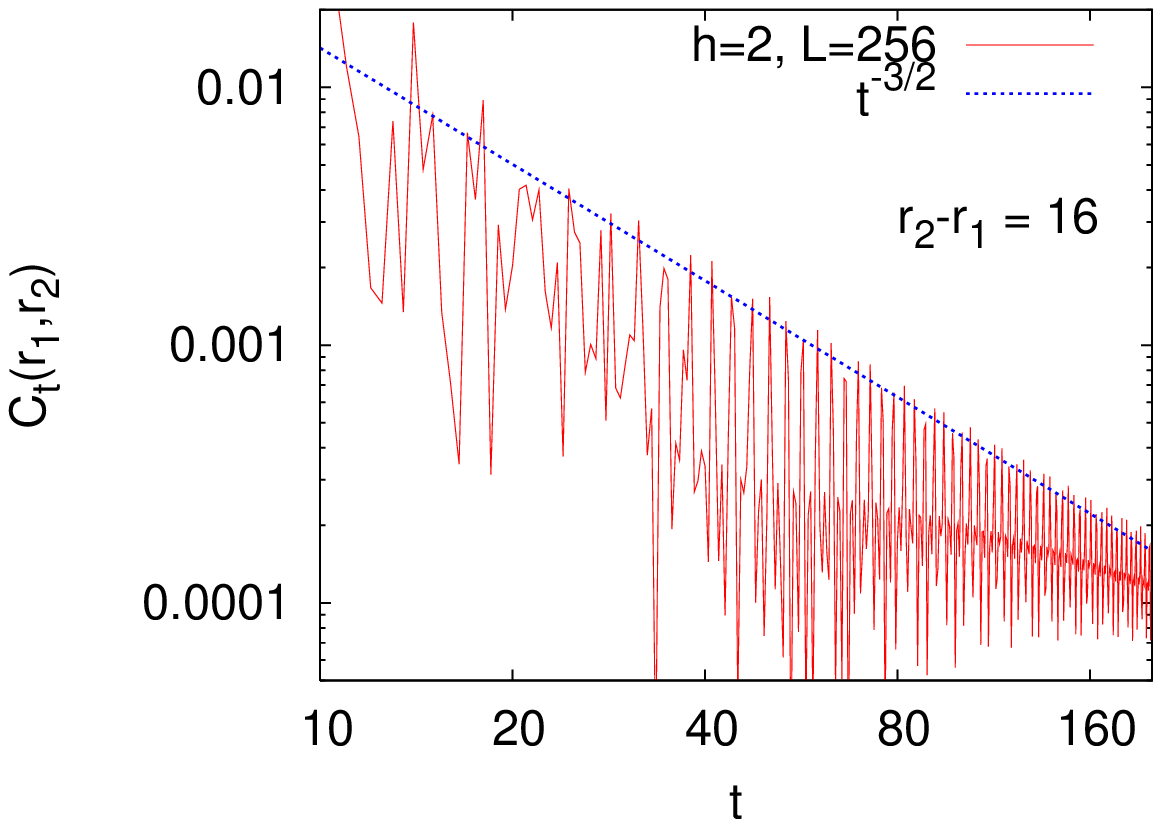}

{\bf C}
\includegraphics[width=6cm]{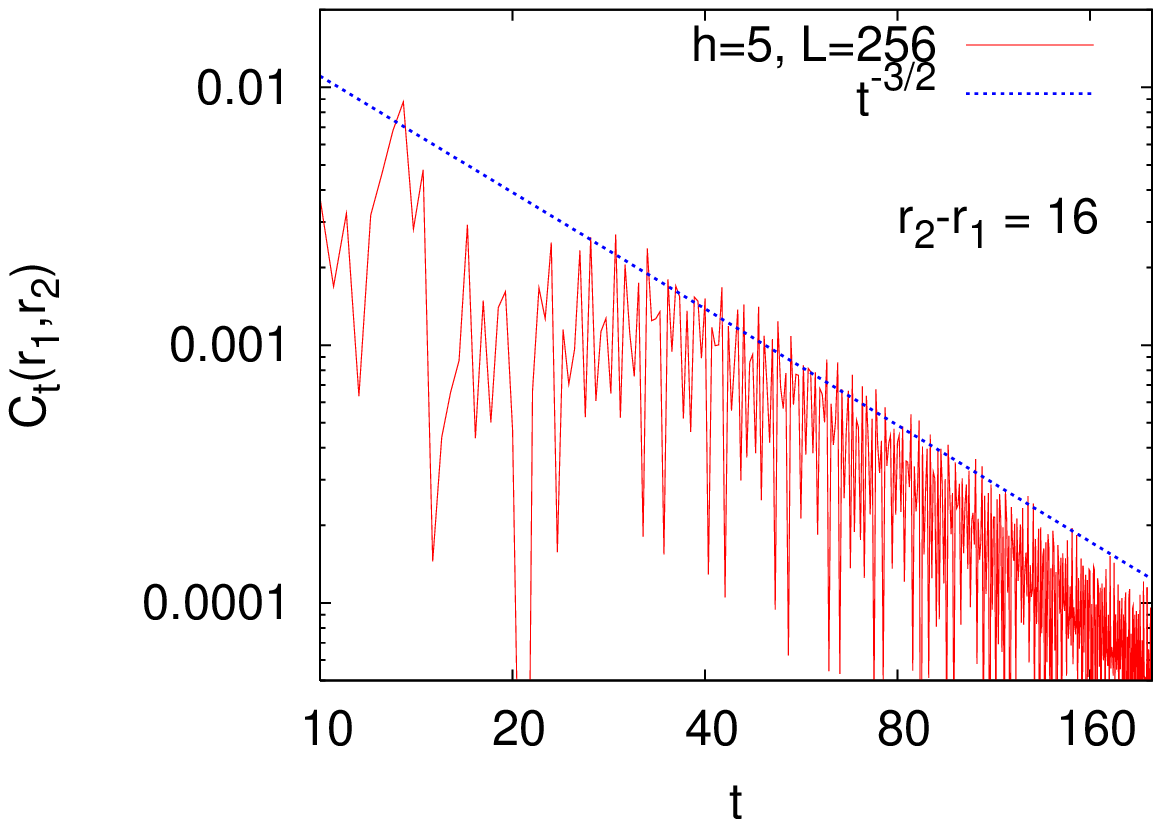}
{\bf D}
\includegraphics[width=6cm]{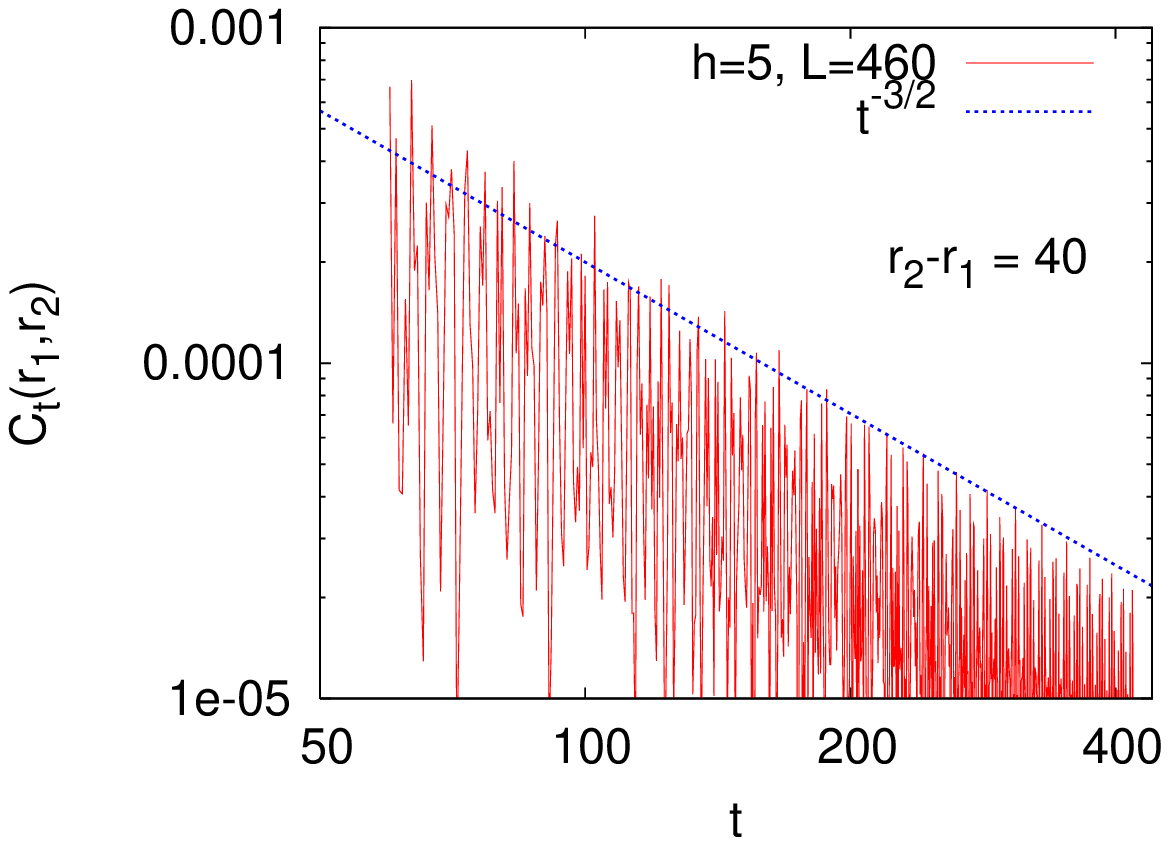}
\caption{
\label{corr-PM} (Color online)
Spatial correlation function $C_t(r_1,r_2)$ as a function of $t$ for
local quenches in the paramagnetic phase ($h>1$) in a log-log plot.
The straight lines correspond to the power law $t^{-3/2}$.
{\bf A:} Different pairs of $(r_1,r_2)$ with the same spatial
difference at $h=3$ ($L$ is 256).
{\bf B:} $h=2$, $L=256$.
{\bf C:} $h=5$, $L=256$.
{\bf D:} $h=5$, $L=460$.
}
\end{center}
\end{figure}

Our results for the spatial correlation function 
$C_t(r_1,r_2)$ within the
paramagnetic phase are depicted in Fig.\ \ref{corr-PM}. We
find that for fixed $r_1$ and $r_2$ an asymptotic power 
law in $t$ with an exponent close to $3/2$:
\be
C_t(r_1,r_2)\propto t^{-3/2}\quad{\rm for}\;t\gg r_1,|r_2|\;.
\ee

\begin{figure}[h!]
\begin{center}
\includegraphics[width=12cm]{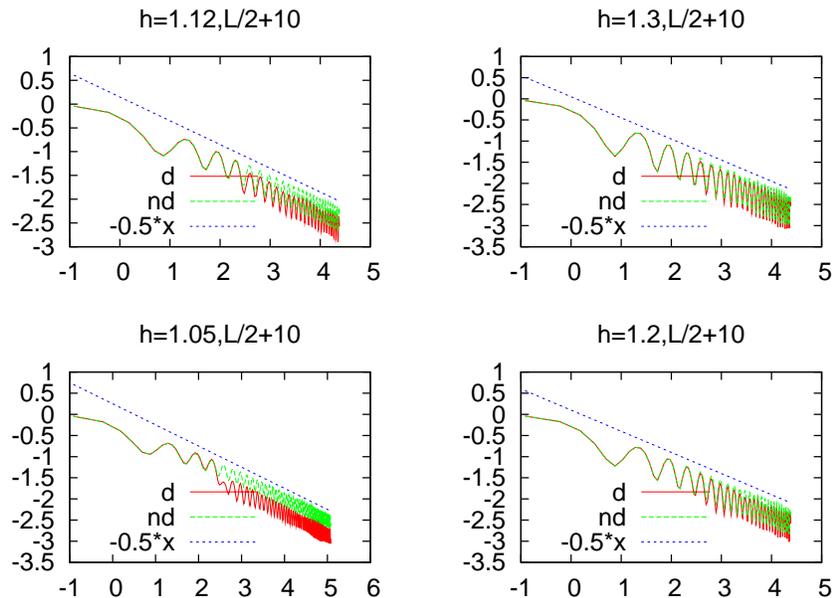}
\caption{
\label{auto-PM} (Color online)
Autocorrelation function 
$G_r(t)$
after a quench in the paramagnetic phase for $r=10$ (i.e.\ for a site in 
a distance 10 from the defect) for different values of $h$ in
comparison with $G_r(t)$ for a system without defect (``nd'').
}
\end{center}
\end{figure}

Finally the autocorrelation function after a local quench in
the paramagnetic phase decays as in the ground state 
without a defect, i.e.\ 
\be
G_t(r)\propto t^{-1/2}\;.
\ee
see Fig.\ \ref{auto-PM}.

\newpage

\section{Conclusions}
\label{sec:concl}
We have studied the temporal evolution of different observables in the transverse Ising
chain following a local quench: for $t<0$ the system consisted of two disconnected halves
which are joined together for $t>0$ with the uniform bulk coupling, $J_{L/2}=1$. We have measured
the magnetization profile, $m_r(t)$, as well as the correlation, $C_t(r_1,r_2)$ and the
autocorrelation function, $G_r(t_1,t_2)$. We have concentrated on the properties of local
quench in the critical state, but some calculations are also performed in the ferromagnetic and
in the paramagnetic phases, too.

For critical local quench several conjectures about
$m_r(t)$ and $C_t(r_1,r_2)$ are known through conformal
field theory, which are valid in the thermodynamic limit and in the continuum approximation.
Our exact finite lattice results have confirmed the conformal conjectures, except for the
early time behavior of the correlation function in which the two reference points are
at different sides of the defect.
We have also studied systematically the finite-size effects, in particular we have made conjectures
about the form of the magnetization profiles, both in space Eq.(\ref{scale2}) and time Eq.(\ref{scale1}),
which probably can be derived in some way, e.g. through conformal field theory.

Our results are explained within the frame of a quasi-particle picture, in which during the quench
kink-like excitations are created at the defect, which move semi-classically, with a momentum-dependent velocity
and result in the reduction of the order-parameter in the system. In the ferromagnetic phase this type
of semi-classical calculation has lead to exact results, at least for a small transverse field, $h$. We
expect, however, that following the same method as in the case of global quench, one can sum the higher order
contributions and in this way one obtains exact results about the stationary value of the magnetization
profile for general value of $h<1$. This involves the non-equilibrium correlation length, $\xi$, as
measured at a global quench, see Eq.(\ref{tilde_m_b}).

Our investigations can be extended and generalized in different directions. First, one can consider
another quantum spin chains, for which the conformal conjectures at the critical point are expected
to be satisfied in the same way as for the transverse Ising chain. In the ferromagnetic phase the
relation in Eq.(\ref{tilde_m_b}) is expected to be valid and in this way one can measure the
non-equilibrium correlation length in an independent procedure. A second way to generalize our results
is to use different forms of the local quench. One possibility is to use a non-zero defect coupling
between the two subsystems in the initial state and/or to have a non-uniform defect coupling, $J_{L/2} \ne J=1$,
in the final state. In the transverse Ising chain the local critical exponents are continuous
function of the strength of the defect \cite{bariev79,McC_Perk,ipt93}, so that $x_s$ for a decoupled system should be
replaced by a defect exponent, $x_d$. Finally, we can also study local defects with a more complicated
structure, which involve several lattice sites.
\bigskip

{\bf Acknowledgments:}\\
One of us (U.D.) is grateful to the Alexander von Humboldt Society for
a fellowship with which this work was performed at the Saarland
University.  This work has furthermore been supported by the Hungarian
National Research Fund under grant No OTKA K62588, K75324 and K77629
and by a German-Hungarian exchange program (DFG-MTA). We thank P. Calabrese
for useful correspondence.

\newpage

{\bf References:}\\

\end{document}